\documentclass[useAMS,usenatbib,twocolumn]{mnras}
\usepackage{natbib}
\usepackage{graphicx}
\usepackage{times}
\usepackage{rotate}
\usepackage{hyperref}
\usepackage{subcaption, tablefootnote}
\usepackage{threeparttable}
\usepackage{amssymb,amsmath,gensymb}
\usepackage{ulem}
\usepackage{url} 

\usepackage{geometry}

\usepackage[usenames, dvipsnames]{color}

\setcitestyle{notesep={; }}

\newcommand{\ds}{\ensuremath{\Delta\Sigma \ }}

\newcommand{\asap}{\texttt{ASAP}}

\newcommand{\later}[1]{\textcolor{magenta}{To be written later. }}

\usepackage{etoolbox}
\makeatletter
\patchcmd\@combinedblfloats{\box\@outputbox}{\unvbox\@outputbox}{}{\errmessage{\noexpand patch failed}}
\makeatother

\begin{document}


\title[The Stellar Profiles of Massive Galaxies: HSC vs. Hydro]{Stellar and Weak Lensing Profiles of Massive Galaxies in the Hyper-Suprime Cam Survey and in Hydrodynamic Simulations}




\author[Ardila et al.]{Felipe Ardila$^{1}$\thanks{E-mail: fardila@ucsc.edu}, Song Huang$^{2}$, Alexie Leauthaud$^{1}$, Benedikt Diemer$^{3}$, Annalisa Pillepich$^{4}$, 
\newauthor Rajdipa Chowdhury$^{1}$, Davide Fiacconi$^{5, 6}$, Jenny Greene$^{2}$, Andrew Hearin$^{7}$, Lars Hernquist$^{8}$, 
\newauthor Piero Madau$^{1}$, Lucio Mayer$^{9}$, S\'ebastien Peirani$^{10, 11}$, and Enia Xhakaj$^{1}$
\\
$^{1}$ Department of Astronomy and Astrophysics, University of California, Santa Cruz, 1156 High Street, Santa Cruz, CA 95064 USA \\
$^{2}$ Department of Astrophysical Sciences, Princeton University, 4 Ivy Lane, Princeton, NJ 08544, USA \\
$^{3}$ Department of Astronomy, University of Maryland, College Park, MD 20742, USA \\
$^{4}$ Max-Planck-Institut f{\"u}r Astronomie, K{\"o}nigstuhl 17, D-69117 Heidelberg, Germany \\
$^{5}$ Institute of Astronomy, University of Cambridge, Madingley Road, Cambridge CB3 0HA, UK \\
$^{6}$ Kavli Institute for Cosmology, University of Cambridge, Madingley Road, Cambridge CB3 0HA, UK \\
$^{7}$ Argonne National Laboratory, Argonne, IL 60439, USA \\
$^{8}$ Harvard-Smithsonian Center for Astrophysics, 60 Garden Street, Cambridge MA 02138, USA \\
$^{9}$ Center for Theoretical Astrophysics and Cosmology, University of Zurich
Winterthurerstrasse 190, CH-8057 Z{\"u}rich, Switzerland \\
$^{10}$ Institut d'Astrophysique de Paris, CNRS \& UPMC, UMR 7095, 98 bis Boulevard Arago, F-75014 Paris, France \\
$^{11}$ Universit\'e C\^ote d'Azur, Observatoire de la C\^ote d'Azur, CNRS, Laboratoire Lagrange, Nice, France
}

\maketitle
\label{firstpage}

 
\begin{abstract} 

 We perform a consistent comparison of the mass and mass profiles of massive ($M_\star > 10^{11.4}M_{\odot}$) central galaxies at $z\sim0.4$ from deep Hyper Suprime-Cam (HSC) observations and from the Illustris, TNG100, and Ponos simulations. Weak lensing measurements from HSC enable measurements at fixed halo mass and provide constraints on the strength and impact of feedback at different halo mass scales. We compare the stellar mass function (SMF) and the Stellar-to-Halo Mass Relation (SHMR) at various radii and show that the radius at which the comparison is performed is important. In general, Illustris and TNG100 display steeper values of $\alpha$ where $M_{\star} \propto M_{\rm vir}^{\alpha}$. These differences are  more pronounced for Illustris than for TNG100 and in the inner rather than outer regions of galaxies. Differences in the inner regions may suggest that TNG100 is too efficient at quenching in-situ star formation at $M_{\rm vir}\simeq10^{13}$ M$_{\odot}$ but not efficient enough at $M_{\rm vir}\simeq10^{14}$ M$_{\odot}$. The outer stellar masses are in excellent agreement with our observations at $M_{\rm vir}\simeq10^{13}$ M$_{\odot}$, but both Illustris and TNG100 display excess outer mass as $M_{\rm vir}\simeq10^{14}$ M$_{\odot}$ (by $\sim$0.25 and $\sim$0.12 dex, respectively). We argue that reducing stellar growth at early times in $M_\star \sim 10^{9-10} M_{\odot}$ galaxies would help to prevent excess ex-situ growth at this mass scale. The Ponos simulations do not implement AGN feedback and display an excess mass of $\sim 0.5$ dex at $r<30$ kpc compared to HSC which is indicative of over-cooling and excess star formation in the central regions. The comparison of the inner profiles of Ponos and HSC suggests that the physical scale over which the central AGN limits star formation is $r\lesssim20$ kpc. Joint comparisons between weak lensing and galaxy stellar profiles are a direct test of whether simulations build and deposit galaxy mass in the correct dark matter halos and thereby provide powerful constraints on the physics of feedback and galaxy growth. Our galaxy and weak lensing profiles are publicly available to facilitate comparisons with other simulations.

\end{abstract}



\section{Introduction}

In the $\Lambda$ cold dark matter ($\Lambda$CDM) cosmological paradigm, dark matter halos (and their associated galaxies) grow hierarchically, building up mass over time through mergers and accretion. However, developing a complete model for galaxy evolution which fully explains the properties of observed galaxies has historically been challenging. For a number of years, massive galaxies in simulations have typically been brighter, bluer, and contained more stellar mass than what is actually observed \citep[e.g.,][]{Borgani&Kravtsov2011:review_cluster_sims}. This has often been referred to as the ``overcooling problem" because it results from excess cooling of gas leading to extended star-formation and overly massive galaxies \citep{Benson+2003}. An energetic feedback process is necessary to quench star formation in massive galaxies and to reproduce the high mass end of the galaxy mass function \citep{Borgani&Kravtsov2011:review_cluster_sims}. Enhanced stellar feedback has been proposed as a solution \citep{Springel&Hernquist:2003a,Springel&Hernquist:2003b}, but recent work has shown that it is simply not enough to account for the necessary energy injection \citep[e.g.][]{Teyssier+2011, Martizzi+2012a}. The most popular scenario for this energy source is feedback by supermassive black holes (BH) at the centers of active galactic nuclei (AGN). 

With the introduction of feedback (from both stars and AGN), many modern hydrodynamic simulations of galaxy formation reproduce, to first order, the observed galaxy stellar mass function \citep[SMF; e.g.][]{Genel+2014:Illustris, Schaye+2015, Beckmann+2017, Pillepich+2018b:TNG}. The subgrid models of hydrodynamical simulations where these feedback processes are numerically approximated are usually calibrated with some set of observables, including: the galaxy SMF, the present-day stellar-to-halo mass relation (SHMR), the star formation rate density (SFRD), the BH mass to galaxy or halo mass relation, the halo gas fraction, galaxy stellar sizes, and the mass-metallicity relation. However, as the demand for even higher fidelity simulations increases, higher order effects must be taken into account. In particular, when calibrating to the galaxy stellar mass function, consistent definitions for the masses of galaxies are not always adopted between the observations and the simulations \citep[e.g., see discussion in][]{Pillepich+2018b:TNG}.

In simulations, stellar mass is the direct outcome of the underlying physical recipe and can be precisely known. Nonetheless, there is still a variety of stellar mass definitions that are adopted, the most common of which are: the sum of all the stellar particles gravitationally bound to a galaxy as defined by some halo finder \citep[e.g.][]{Remus+2017, Dubois+2016}, the sum of all stellar particles within 3D spherical apertures at fixed radius \citep[e.g.][]{Schaye+2015}, or some combination of those two definitions \citep[e.g.][]{Pillepich+2018b:TNG}. Other works have also created mock data from the simulations and performed observationally-motivated measurements in order to recover the mass in a way that is more consistent with observations \citep[e.g.][]{Price2017, Laigle+2019}.

In data, stellar mass estimates depend on the mass-to-light ratio estimates (M/L), modeling of galaxy light profiles to extract galaxy luminosity (L), sky subtraction, knowledge of redshifts, and can also depend galaxy morphology \citep[e.g.][]{Bernardi+2013, D'Souza+2015, Muzzin+2013}. In this paper, however, we focus specifically on very massive galaxies ($M_\star > 10^{11.4}M_{\odot}$), which present different challenges compared to the general galaxy population. Super massive galaxies have redshifts that are generally well known. They are red and form a fairly homogeneous population with little dust and relatively shallow (M/L) gradients. Instead, the dominant challenge for super massive galaxies is the estimate of their luminosity due to their extended diffuse components that extend to 100 kpc and beyond. These outer low surface brightness regions of a galaxy can lead to important corrections in stellar mass \citep{Bernardi+2013, D'Souza+2015, Huang+2018b}. However, recently, deep and wide multi-band imaging from the Subaru Strategic Program \citep[SSP;][]{Aihara+2017:SSP, Aihara+2018} using Hyper Suprime-Cam \citep[HSC;][]{Miyazaki+2012:HSC, Miyazaki+2018} has advanced our understanding of the light profiles of individual very massive galaxies out to 100 kpc and beyond \citep{Huang+2018b, Huang+2018c}. This allows for a robust determination of the stellar mass within 100 kpc ($M_{\star}^{100}$) as well as the full shape of the galaxy light profile from 10 to 100 kpc.

The goal of this paper is to outline a framework for performing a more direct comparison between observations and hydrodynamic simulations that capitalizes on new generation surveys that are wide, deep, and that have weak lensing capabilities. First, with surveys such as HSC, weak lensing enables comparisons at fixed halo mass \citep[e.g.,][]{Huang+2019}. Second, in addition to making use of weak lensing, we also advocate for the comparison of galaxy mass profiles. This avoids having to define ``galaxy mass" as a single number and can account for the fact that observations have finite depth. Finally, the radial shape of the galaxy mass profile also contains information about the assembly history and feedback processes that have shaped massive galaxies.

This paper is organized as follows: in \S \ref{sec:data} we describe our data for both observations and simulations; we then discuss how we measure mass density profiles in observations (\S \ref{sec:measurements:data}) and simulations (\S \ref{sec:measurements:sims}); we present results from comparing measurements in observed and simulated galaxies in \S \ref{sec:results} and discuss potential reasons for disagreement in \S \ref{sec:discussion}; finally we summarize and conclude in \S \ref{sec:conclusions}. We provide median surface mass density profiles and weak-lensing $\Delta\Sigma$ profiles for the HSC galaxies here: \url{https://github.com/f-ardila/HSC_vs_hydro-paper}.

The following cosmological parameters are assumed throughout: $H_{0}=70 \ \mathrm{km} \ \mathrm{s}^{-1} \ \mathrm{Mpc}^{-1}$, $\Omega_{\mathrm{m}}=0.3$, and $\Omega_{\Lambda}=0.7$. We use the virial mass for dark matter halo mass ($M_{\mathrm{vir}} = 4\pi \Delta(z) \rho_{\mathrm{c}}(z) R_{\mathrm{vir}}^{3} / 3$) as defined in \citet{Bryan&Norman1998}. All length units use physical units (not comoving).

\section{Data and Simulations}\label{sec:data}

\subsection{HSC data}
In this paper, we take advantage of new high quality imaging from the WIDE layer of the HSC SSP, which is a simultaneously wide ($>1000$ deg$^2$) and deep (r $\sim 26$ mag) imaging survey \citep[with the DEEP and ULTRADEEP layers reaching r $\sim 27$ and r $\sim 28$ and a coverage of 27 deg$^2$ and 3.5 deg$^2$ respectively;][]{Aihara+2017:SSP, Aihara+2018}, meaning that we can observe a large sample of massive galaxies and detect their faint extended stellar envelopes. The imaging is deep enough that we are able to measure surface brightness profiles of individual massive galaxies to $\sim28.5$ mag arcsec$^{-2}$ in \textit{i}-band \citep{Huang+2018b}. 

This work uses the internal s16A data release (equivalent to ``DR1") which covers $\sim 140\ \rm{deg}^2$ in all five broad-bands (grizy) to full WIDE depth. The combination of a wide (1.5$\degree$) field of view and exceptional imaging depth and quality (median i-band seeing FWHM $\sim 0.6$'') makes this survey ideally suited to study the surface brightness profiles of galaxies out to large radii. HSC $i$-band images were used to make our surface brightness (and stellar mass density) profile measurements because of the superior seeing in this band as a result of strict requirements imposed by weak-lensing science.

We use the HSC massive galaxy sample from \citet{Huang+2018b} and refer the reader to that paper for the full details regarding the construction of this sample; we give only a brief summary here. Our sample consists of $\sim 15000$ galaxies with reliable spectroscopic redshifts in the range $0.3 \leq z \leq 0.5$. Our selection begins with an initial magnitude cut of $i_{\mathrm{HSC}, \text { cModel }} \leq 21.5$~mag to select massive galaxies ($\log \left(M_{\star} / M_{\odot}\right) \geq 11.5$) at $z<0.5$ based on \citet{Leauthaud+2016}. We limit our sample to regions that have full depth coverage in $i$-band in the WIDE layer, objects without deblending errors, with well-defined centroids, and usable cModel magnitudes in all five HSC bands. We also exclude objects affected by pixel saturation, cosmic rays, or other optical artefacts. 

We then include only objects with a reliable spectroscopic redshift and restrict redshifts to the range $0.3 \leq z \leq 0.5$. We focus on this range for several reasons: 1) to resolve and reliably measure the stellar mass within the inner 10 kpc of galaxies; 2) to limit background noise and cosmological dimming when measuring stellar mass out to 100 kpc and beyond; 3) to ignore the redshift evolution of the stellar populations in these galaxies; 4) to ensure stellar mass completeness of our sample; 5) to reduce the known issue of over-subtraction of the background that occurs at lower redshifts. \citet{Huang+2018b} have shown that the light profiles of individual galaxies in this sample can be mapped to 100 kpc and beyond (without using stacking techniques).

\subsection{Illustris and TNG100 simulation data}\label{sec:sims}

In this work we focus on two uniform-volume (i.e. not zoom-in) cosmological hydrodynamic simulations: the Illustris-1 simulation of the Illustris Project \citep[referred to hereafter as ``Illustris'';][]{Vogelsberger+2014:Illustris, Vogelsberger+2014:Nature, Genel+2014:Illustris, Sijacki+2015:Illustris} and the TNG100 simulation of the IllustrisTNG Project \citep[referred to hereafter as ``TNG100'';][]{Springel+2018:TNG, Marinacci+2018:TNG, Pillepich+2018b:TNG, Naiman+2018:TNG, Nelson+2018:TNG}. Table \ref{tab:sims} gives the main parameters of both simulations. They are also briefly described below.

Illustris consists of three full physics simulation boxes of the same size ($75\; h^{-1}$Mpc), but varying resolutions, run to $z=0$ with the moving-mesh \texttt{AREPO} code. There are also additional runs with dark matter only and adiabatic scenarios. For this project we use the highest resolution, full physics run (Illustris-1). The Illustris galaxy formation model includes gravitational interactions among the different resolution elements (dark matter, stars, gas, and black holes); hydrodynamical equations for the gas component; a treatment of radiative cooling and heating processes; a mechanism for converting gas into stars; stellar evolution and the resulting chemical enrichment of the interstellar, circumgalactic and intergalactic media; stellar feedback induced outflows of gas; and the formation, growth and energetic feedback of supermassive black holes in distinct low- and high-accretion rate states \citep{Vogelsberger+2013, Torrey+2014}. The fiducial model was chosen to broadly reproduce the $z=0$ galaxy stellar mass function, the evolving cosmic SFR density, and the $z=0$ relation between galaxy mass and gas-phase metallicity \citep{Torrey+2014}.

IllustrisTNG (``The Next Generation'') is the successor to Illustris and has a larger range of box sizes and resolutions and several enhancements. For this project we use the highest resolution of the medium size TNG100 run, which has the same $75 \ h^{-1}$Mpc box size and similar resolution as Illustris-1. The IllustrisTNG runs include improved numerical methods, new physics (e.g. magnetohydrodynamics), and modified sub-grid physical models to address the key shortcomings of the Illustris model. Most importantly, the new dual mode (thermal and kinetic) AGN feedback model helps to better regulate the stellar content of massive galaxies while preserving realistic halo gas fractions; and the galactic winds feedback model has been updated to better reproduce the abundance or mass of intermediate- and low-mass galaxies \citep{Pillepich+2018a, Weinberger+2017}. The fiducial IllustrisTNG model was chosen by comparing the outcome of many TNG model variations to that of the original Illustris model (and not directly to observations) and by ultimately choosing an implementation and parameter set that simultaneously had the potential to alleviate, at least qualitatively, the largest number of targeted tensions between Illustris and observations. For this purpose, the comparison focused on the global SFR density as a function of time; the current galaxy stellar mass function; the stellar mass-halo mass relation at $z=0$ in addition to the BH mass versus stellar or halo mass relation; the gas fraction within the virial radii of haloes; and the stellar half-mass radii of galaxies.

Galaxy stellar masses between Illustris and IllustrisTNG were compared by adopting the same operational definition, namely the sum of gravitationally bound stellar particles contained within twice the stellar half-mass radius of each SUBFIND (sub)halo. This was also the operational definition used for the original ``calibration'' of the Illustris model, against SMFs measured by \citet{Baldry+2008} and \citet{Perez-Gonzalez+2008}. It should be kept in mind, however, that in the development of the Illustris and IllustrisTNG frameworks, the best model parameter values have {\it not} been obtained via an actual fit to a selection of galaxy observables, but simply by requiring consistent trends and overall normalizations. The adopted approach was intentionally kept simple because an actual fine tuning of galaxy population models against observational data would have required complex steps to be applied to the output of hundreds of model variations (e.g., transforming raw simulated data into realistic mock observations and applying selection functions tailored to each targeted observational data-set). Furthermore, the very high-mass end of the galaxy population could not be taken into account during the development phase, because of the computational costs of simulating hundreds of large volumes at fixed resolution. The high mass end is therefore a regime where the models are de facto predictive.

For this paper, we choose all central galaxies in Illustris and TNG100 with total stellar masses $M_* > 10^{11.2} M_{\odot}$ at $z = 0.4$ (snapshot 108 in Illustris and 72 in TNG). This results in a sample of 339 galaxies in Illustris and 235 galaxies in TNG100. We choose this redshift in order to be consistent with the median redshift of our observations. Here, a central galaxy is defined by the \textsc{Subfind} halo finder as the most bound subhalo within a larger friends-of-friends (FOF) group. All other bound halos within the FOF group (and their galaxies) will be denoted as subhalos and satellites, respectively. This stellar mass cut results in a $\sim85\%$ ($\sim91\%$) completeness of central galaxies with M$_{\mathrm{vir}}>10^{13} \ M_{\odot}$ for TNG100 (Illustris).

\begin{table*}
\begin{threeparttable}
	\centering
	\begin{tabular}{lccccccccccc} 
		\hline
		Simulation & Size & N$_{\mathrm{particles}}$ & $m_{\mathrm{DM}} $ & $m_{\mathrm{baryon}}$ & $\epsilon_{\mathrm{DM}} $ & $\epsilon_{\mathrm{gas, min}}$  & $\Omega_{\mathrm{M}}$ & $h$ &
		N$_{\mathrm{halos}}\tnote{*}$ & hydrodynamics & Reference \\
          & $[Mpc]$ &  & $[M_\odot]$ & $[M_\odot]$ & $[kpc]$ & $[kpc]$ & && & code  \\
		\hline\hline
        Illustris & 106.5 & $2 \times 1820^3$ & $6.3 \times 10^6$ & $1.3 \times 10^6$ & 1.42 & 0.71 & 0.2726 & 0.704 &160& \texttt{AREPO} & \citet[][]{Nelson+2015:illustris}\\
        TNG100 & 110.7 & $2 \times 1820^3$ & $7.5 \times 10^6$ & $1.4 \times 10^6$ & 0.74 & 0.185 & 0.3089 & 0.6774 &214&\texttt{AREPO} & \citet[][]{Nelson+2019:tng}\\
        \hline
        Ponos & -- & $2 \times 10^6$ & $2.3 \times 10^6$ & $4.5 \times 10^5$ & 0.785 & 0.210 & 0.272  & 0.702 &--& \texttt{Gasoline} &  \citet{Fiacconi+2017}\\
		\hline
	\end{tabular}
    \begin{tablenotes}
    
        \item[*] $M_{\mathrm{vir}}>10^{13}M_{\odot}$
    \end{tablenotes}
    \caption{Comparison of the details of the various hydrodynamic simulations.}
	\label{tab:sims}

\end{threeparttable}
\end{table*}

\subsection{Ponos simlation data}\label{sec:ponos_data}
We also compare the stellar mass profiles of our HSC galaxies to two highly resolved galaxies from the Ponos zoom-in numerical simulations of dark matter substructures in massive ellipticals \citep{Fiacconi+2016, Fiacconi+2017} including hydrodynamics using the TreeSPH code \texttt{GASOLINE2} \citep{Wadsley+2004}, which employs a modern implementation of the SPH equations based on the geometric density formulation of the pressure
force, sub-grid turbulent diffusion of thermal energy and metals, and the Wedland C4 kernel. At $z=0.4$, the two Ponos galaxies (PonosV and PonosSB) have identical dark matter halo masses ($M_{\mathrm{\rm vir}}=1.04 \times 10^{13} M_{\odot}$) but different stellar feedback mechanisms resulting in slightly different stellar masses between the two ($M_{\star}^{\mathrm{100}}\sim 10^{12} M_{\odot}$). 

PonosV\footnote{Also referred to as `PonosHydro’ or `PH' in \citet{Fiacconi+2017} to distinguish it from the DM-only counterpart in \citet{Fiacconi+2016}.} (`V' for violent merger history compared to more quiescent halos) adopts the blastwave feedback sub-grid model \citep{Stinson+2006}, in which thermal feedback from explosions of SN Type II is achieved by locally shutting off cooling for a timescale comparable to the duration of the Sedov-Taylor and snowplough phases alltogether. Blastwave feedback can lead to realistic stellar masses and structural properties of disk galaxies provided that the internal parameters are set according to resolution \citep{Sokolowska+2017}. PonosSB employs the more recent Superbubble (SB) feedback implementation by \citet{Keller+2014}, in which thermal energy and ejecta are deposited into multiphase particles, with the warm/hot phase returning to the cold phase over a few Myr owing to thermal conduction. Superbubble feedback does not require to shut-off cooling, and has been shown to be more efficient at regulating stellar masses in low mass and disk galaxies (up to $M_{\mathrm{vir}} \sim 10^{12}$) in a resolution-independent way, matching easily the stellar mass-to-halo mass relation in this mass range \citep{Keller+2015}. 

Internal parameters of the two sub-grid feedback models, and of the underlying star formation recipe, are set as in \citet{Mayer+2016}. Neither of the two galaxies include AGN feedback, and both include radiative cooling from metal lines. Comparisons with the mass profiles from these Ponos galaxies and Illustris/TNG100 will shed light both on the impact of the resolution of the simulations, as well as the impact of AGN feedback \citep[but see also Appendices and discussions on the effects of both resolution and AGN feedback within the TNG model itself in][]{Pillepich+2018a,Pillepich+2018b:TNG}

		

\subsection{Mass maps}\label{sec:mass_maps}

One of the main goals of this paper is to perform a consistent comparison between observations and theory. To treat simulated galaxies in a similar way as observations, we first translate the 3D distribution of their stellar population particles into 2D stellar mass maps. We choose to use projected mass maps rather than mock multi-band observations for each galaxy because we wish to limit the number of assumptions we need to make. Because we are interested in studying mass profiles, mock observations would require additional assumptions to convert these to stellar mass. Even if we were to study luminosity profiles rather than mass, we would still need assumptions about stellar population models to perform the k-corrections to build average luminosity profiles of our observed sample. Therefore, we choose to put these assumptions only on the observational side, and keep the simulation data as is.

For each galaxy, we create a map by projecting the positions of stellar population particles onto a $300^2$ grid of pixels. We have experimented with smoothing the maps using a Gaussian kernel, but find that such smoothing can lead to slight errors in the integrated stellar mass in the map. Thus, we simply assign the total mass of a stellar particle to the map pixel in which its center lies.

Our maps have a pixel size of $1.0$ kpc/pixel and a physical side length of $300$ kpc. These allow us to more accurately trace galaxy profiles on the scales that we are interested, below $r<100$ kpc. We tested different resolution scales and found that with maps of $1.0$ kpc/pixel, the measured mass profile was stable beyond $6$ kpc. Within $6$ kpc, we are limited by seeing in observations anyway, so we do not compare profiles below this scale.

We use all stellar population particles within the FOF group (i.e. we impose the particular projection depth determined by the FOF) to perform various tests. We distinguish between particles bound to the galaxy, particles bound to satellite galaxies in the same group, and particles that are not bound to any galaxy in the group. In practice, the latter component is negligible, meaning that all the stars in our maps are bound to either the central galaxy or its satellites. We also find the satellite component does not affect the measured mass profile within 100 kpc (\S\ref{sec:masking_sats}) so we only include mass maps with the central galaxies in our analyses.

\section{Measuring Masses and Mass Profiles in Data} \label{sec:measurements:data}
\subsection{One-Dimensional Surface Mass Density Profiles}\label{sec:profiles_in_data}

We measure the 1D stellar mass density profiles of HSC galaxies from their HSC $i$--band images using the method of \citet{Huang+2018b}. We use the galaxy surface brightness profile function (\texttt{galsbp}) in the publicly available \texttt{kungpao} package\footnote{\url{https://github.com/dr-guangtou/kungpao}}. First, we apply an empirical background correction and mask out neighboring objects based on their brightness and proximity. Then, we draw concentric elliptical isophotes with a fixed ellipticity on the target, and at a given radius (along the semi-major axis), we measure the median intensity along each isophote after iterative 3$\sigma$--clipping. The ellipticity parameters are computed from the intensity-weighted moments of
flux distribution of the object in the map using \texttt{SEXTRACTOR} \citep[][through the \texttt{SEP PYTHON} library]{sextractor}. Hereafter, in all figures, R corresponds to a distance along the semi-major axis of the elliptical isophote. 

We then convert these profiles into surface stellar mass profiles ($\mu_*$) by assuming a radially constant mass-to-light ratio measured from SED fitting. We note that massive elliptical galaxies are known to have shallow negative color gradients \citep[e.g.][]{Carollo+1993, Davies+1993, LaBarbera+2012, D'Souza+2015}, which may in principle underestimate the stellar mass in the centre, while overestimating the stellar mass in the outskirts. However, because the gradients are shallow, and they are smooth out to a few times the effective radius \citep[e.g.][]{LaBarbera+2010, D'Souza+2014}, an average M/L is unlikely to bias our stellar mass measurements \citep[for details see Appendix C of][]{Huang+2018b}.

Integrating $\mu_*$ profiles provides us with the stellar mass ($M_*$) within elliptical apertures. We use the stellar mass within a 100 kpc aperture (hereafter noted $M_{\star}^{\mathrm{100}}$) as a value close to the total stellar mass of a galaxy. We also use the stellar mass within a 10 kpc aperture ($M_{\star}^{\mathrm{10}}$) as the inner stellar mass of the galaxy, and the difference ($M_{\star}^{\mathrm{100}} - M_{\star}^{\mathrm{10}}$) as the outer stellar mass. Estimates for exactly how much mass is missed by cutting off at 100 kpc will be given in a follow-up paper (Ardila et al. in preparation). As shown in \citet{Huang+2018b} and Li et al. in preparation, at large scales we can reliably measure $\mu_*$ profiles for individual galaxies out to more than 100 kpc without being limited by the background subtraction, while on small scales our profiles are resolved down to $\sim 6$ kpc. As a reference, 1.0$^{''}$ corresponds to 3 and 6 kpc at $z=0.2$ and $0.5$, respectively, and the mean $i$-band seeing has FWHM $=0.58^{''}$. 

Our \texttt{HSC WIDE} data can reach $>29$ mag$/$arcsec$^{2}$ in surface brightness radial profile measurements in the $i$-band. It is known that the \texttt{hscPipe} deblending process is not optimized for these extended objects and that it tends to oversubtract background light. To avoid these problems we use \texttt{SEXTRACTOR} \citep{sextractor} background subtraction and object detection (using the \texttt{SEP PYTHON} library) to generate the appropriate masks. Appendix B in \citet{Huang+2018b} gives more details about the masking performed in our HSC sample. As is shown in that paper, different masking methods do not affect the measured surface brightness profiles of HSC galaxies within 100 kpc. Even past 100 kpc, the difference resulting from different masking methods is small. We further show that the effect of satellite contamination is negligible with our simulated galaxies in \S \ref{sec:masking_sats} (see also Figure \ref{fig:masking_satellites}). 

We are unable to measure reliable 1D profiles for $\sim 11 \%$ of the massive galaxies in our sample due to a complex inner structure (e.g. on-going a major merger), or substantial contamination from a bright star or foreground galaxy, but excluding them does not bias our sample, nor does it significantly affect the measured SMF \citep{Huang+2018b}.

\subsection{Halo Masses from Weak Lensing} \label{sec:measurements:data:weak-lensing}

Thanks to the unprecedented weak lensing capability of the HSC survey, we are able to perform galaxy-galaxy lensing (g-g lensing hereafter) measurements around a large sample of nearby massive galaxies, and infer their halo mass through careful galaxy-halo connection modeling (e.g. 
\citealt{Huang+2019}). g-g lensing measures the coherent distortions in the shapes of background galaxies due to the mass of a foreground lens galaxy and its dark matter halo. In practice, we are measuring the excess surface mass density profile ($\Delta\Sigma$) defined as

\begin{equation}
\Delta\Sigma\left(r_{p}\right) = 
\overline{\Sigma}\left(< r_{p}\right)-\Sigma\left(r_{p}\right)=\gamma_{t}\left(r_{p}\right)\Sigma_{\rm crit},
\label{eq:ds1}
\end{equation}

\noindent where $\overline{\Sigma}\left(< r_{p}\right)$ is the mean projected surface mass density within radius $r$ and $\Sigma\left(r_{p}\right)$ is the azimuthally averaged surface mass density at radius $r$. 
$\gamma_{t}\left(r_{p}\right)$ is the tangential shear component, and $\Sigma_{\rm crit}$ is the critical surface density. 

For our massive galaxies, we measure their $\Delta\Sigma$ profiles using the weak lensing shape catalog defined by \citet{Mandelbaum+2018}. We follow the strategy outlined by \citet{Singh+2017} and use the \texttt{dsigma} code \footnote{\url{https://github.com/dr-guangtou/dsigma}}, which is optimized to work with HSC-SSP data. See \citet{Speagle+2019} and \citet{Huang+2019} for details of the g-g lensing measurements.

With the help of these g-g lensing profiles and the deep stellar mass density profiles, we are able to build an empirical model that connects the stellar mass distribution of massive galaxies to their underlying halo masses (\citealt{Huang+2019}). This model (referred to as \texttt{ASAP}) is based on the Small MultiDark Planck \citep[SMDPL;][]{Klypin+2016:MDPL2} N-body simulation of halo merging history and a modified version of the \texttt{UniverseMachine} semi-empirical model (\citealt{Behroozi+2019}). \texttt{ASAP} is constrained by weak lensing data and by HSC observations of massive galaxies with $M_{\star}^{\mathrm{100}}>10^{11.6} M_{\odot}$. In this mass range, the satellite fraction is very low ($<10$\%). \texttt{ASAP} provides the link between the halo masses ($M_{\mathrm{vir}}$) of massive central galaxies, the stellar mass within 10 kpc ($M_{\star}^{\mathrm{10}}$), and the ``total'' stellar mass ($M_{\star}^{\mathrm{100}}$)\footnote{In practice, \asap~ is modeled using $M_{\star}^{\mathrm{1D,Max}}$, which is the maximum stellar mass one can measure from 1-D surface brightness profile without any extrapolation, rather than $M_{\star}^{\mathrm{100}}$. For the $M_{\star}^{\mathrm{100}}>10^{11.6}$ HSC sample, the mean difference between $\log(M_{\star}^{\mathrm{1D,Max}})$ and $\log(M_{\star}^{\mathrm{100}})$ is $\sim0.02$ dex, so we choose to use $M_{\star}^{\mathrm{100}}$ in the notation of this paper for simplicity.}. The intrinsic scatter in this relation is 0.15-0.20 dex. 

\subsection{Comparisons at Fixed Halo Mass} \label{sec:measurements:data:fixedmhalo}

In this work, we compare observed galaxies to their simulated counterparts at fixed halo mass. For real HSC galaxies  we can only directly measure $M_{\star}^{\mathrm{10}}$ and $M_{\star}^{\mathrm{100}}$ so we can only know their mean halo masses through a calibrated scaling relation. Given the $M_{\star}^{\mathrm{10}}$ and $M_{\star}^{\mathrm{100}}$ values for HSC galaxies we assign mean halo mass values following the \texttt{ASAP} model. This relation is given by:

\begin{align}\label{eq:meanmhalo}
    \log \overline{M}_{\rm vir} =& 3.26 \times (\log M_{\star}^{\mathrm{100}} - 11.72) \\
    \nonumber -& 2.46 \times (\log M_{\star}^{\mathrm{10}} - 11.34) \\
    \nonumber+& 13.69
\end{align}

\noindent Halo masses assigned in this fashion will be noted hereafter, $\overline{M}_{\rm vir}$. We adopt Equation \ref{eq:meanmhalo} when we select individual HSC galaxies for comparisons in bins of halo mass (e.g. in \S \ref{sec:match_by_m_halo}).

Given the \asap\ model that best describes the HSC data \citep[][]{Huang+2019}, we can also calculate the full halo mass distribution, $P(M_{\rm vir}|\overline{M}_{\rm vir})$, corresponding to any given cut in $\overline{M}_{\rm vir}$. In practice, we measure $P(M_{\rm vir}|\overline{M}_{\rm vir})$ from the full distribution of halo masses in the SMDPL simulation which is populated with galaxies drawn from the best-fit \asap\ model. Additionally, for a given $\overline{M}_{\rm vir}$ cut in HSC, we can select a distribution of halos in Illustris/TNG that approximately matches $P(M_{\rm vir}|\overline{M}_{\rm vir})$. Note that this matching relies on modeling the HSC data with the ASAP model. However, the ASAP model is more sophisticated than most models used in past work (e.g. Halo Occupation Models) and was specifically designed to interpret lensing data for super massive galaxies. In the text, we will clarify when $\overline{M}_{\rm vir}$ and $P(M_{\rm vir}|\overline{M}_{\rm vir})$ are used.

\section{Measuring Masses and Mass Profiles in Simulations} \label{sec:measurements:sims}

\subsection{Mass profiles in simulations}

To measure the mass profiles of simulated galaxies, we adopt a nearly identical procedure as described in \S \ref{sec:profiles_in_data} for observed galaxies. The only difference is that we start with 2D projected stellar mass maps of each galaxy and measure the surface mass density profiles directly from those (i.e. we do not need to consider the mass-to-light ratio in the simulations). As with the HSC data, we draw concentric elliptical isophotes with a fixed ellipticity on each galaxy, and measure the mean intensity at a given radius along the semi-major axis of each isophote, after applying iterative 3$\sigma$--clipping. The ellipticity and position angle of the isophotes are set to the average values from fitting the 2D shape of each galaxy. The techniques we applied to HSC data to subtract contributions from backgrounds and foregrounds are not required by our simulation analysis since these contaminants are not present in the stellar particle data extracted from the simulation output based on halo finders, as used here. This allows us to test the impact of masking satellites in the next section (\S \ref{sec:masking_sats}). 

\subsection{Testing the Effect of Masking Satellites} \label{sec:masking_sats}

In order to reliably measure the surface brightness profiles of central galaxies, it is necessary to mask out nearby satellites, foreground, and background sources. This can be challenging at the depth of HSC, especially for our sample of massive ellipticals with very extended profiles and for which nearby satellites are common. As mentioned in \S \ref{sec:profiles_in_data} \citep[details can be found in][]{Huang+2018b} different masking choices did not affect the measured surface brightness profile of HSC galaxies within 100 kpc.

\begin{figure}
      \centering 
      \includegraphics[width=8cm]{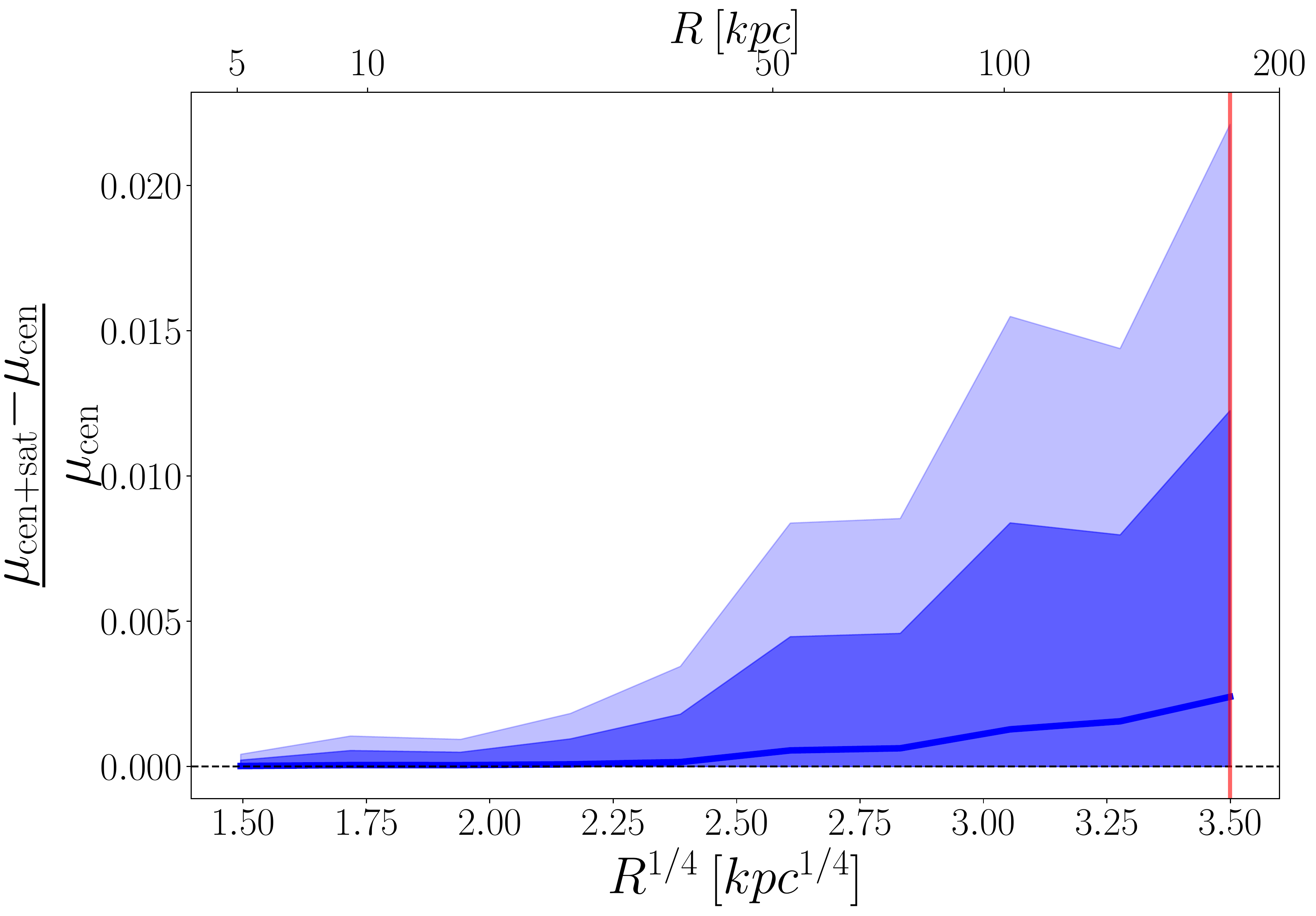}
      \caption{Mean fractional difference in the surface mass density profile extracted by applying the same 3$\sigma$--clipping method for satellite removal to high resolution maps with both centrals and satellites ($\mu_{\mathrm{cen+sat}}$), and high resolution maps with only the central galaxy ($\mu_{\mathrm{cen}}$) for $z=0.4$ TNG100 galaxies in our sample ($M_{\star}>11.2$). The shaded regions show the 1$\sigma$ and 2$\sigma$ offsets. The figure shows that on average there is no difference when the maps for our measurements include satellites, until $\sim 50$ kpc, at wich point the difference begins to grow and reaches $\sim 0.02\%$ at 150 kpc (the size of our high resolution maps indicated by the red vertical line). The 1$\sigma$ (2$\sigma$) region is maximum at 150 kpc with a $\sim 1\%$ ($\sim 2\%$) effect. Satellites do not affect our surface mass density measurements because they are generally excluded by the 3$\sigma$--clipping of our method (\S \ref{sec:masking_sats}). }
      \label{fig:masking_satellites}
  \end{figure}

Our simulated galaxies also allow us test the effects of masking satellites in the surface brightness profile measurements of HSC data. Given that we can easily include or exclude particles belonging to satellites, we can measure the effect satellites have on the surface mass density profile. Figure \ref{fig:masking_satellites} shows the mean, 1$\sigma$, and 2$\sigma$ fractional difference in the mass density profiles between mass maps with and without satellites, when the same 3$\sigma$--clipping procedure to remove satellite contamination is applied. Even at the 2$\sigma$ level, when including satellites (i.e. without any masking applied at all) the difference in the mass density profile will be less than $2\%$ within 150 kpc. Since this test only compares perfect masking to non-masking, any realistic satellite masking performed on our data can be expected to perform better than the non-masking situation. This demonstrates that masking is not so important when using our profile measurement technique within 100 kpc.

 While it has been shown that the stellar mass in satellites is not negligible compared to the total stellar mass of a cluster \citep[e.g.][]{Pillepich+2018b:TNG}, our results show that including satellites will not heavily affect the measurement of the mass density profile of a galaxy. This can be understood through the technique used to make the measurement. As described in \S \ref{sec:profiles_in_data}, along each isophote we measure the mean intensity \textit{after} iterative 3$\sigma$--clipping. This means that any satellites along an isophote that would heavily bias our measurement due to their high mass intensity would be excluded by the 3$\sigma$--clipping. These results shows that the details of our masking strategy do not impact our HSC results, and also that our results are not affected by satellites below our detection threshold. 

It is worth noting that this test only includes physically associated satellite galaxies. In reality, surface brightness profiles may also be contaminated by nearby foreground and background objects. Using mock galaxies on a more recent HSC data release, Li et al. (in prep.) uses a more realistic scenario to further demonstrate that our masking strategy can secure the surface brightness profile out to 100 kpc.

\section{Results}\label{sec:results}
\begin{figure*}[htb]
  \centering 
    \includegraphics[width=0.35\textwidth]{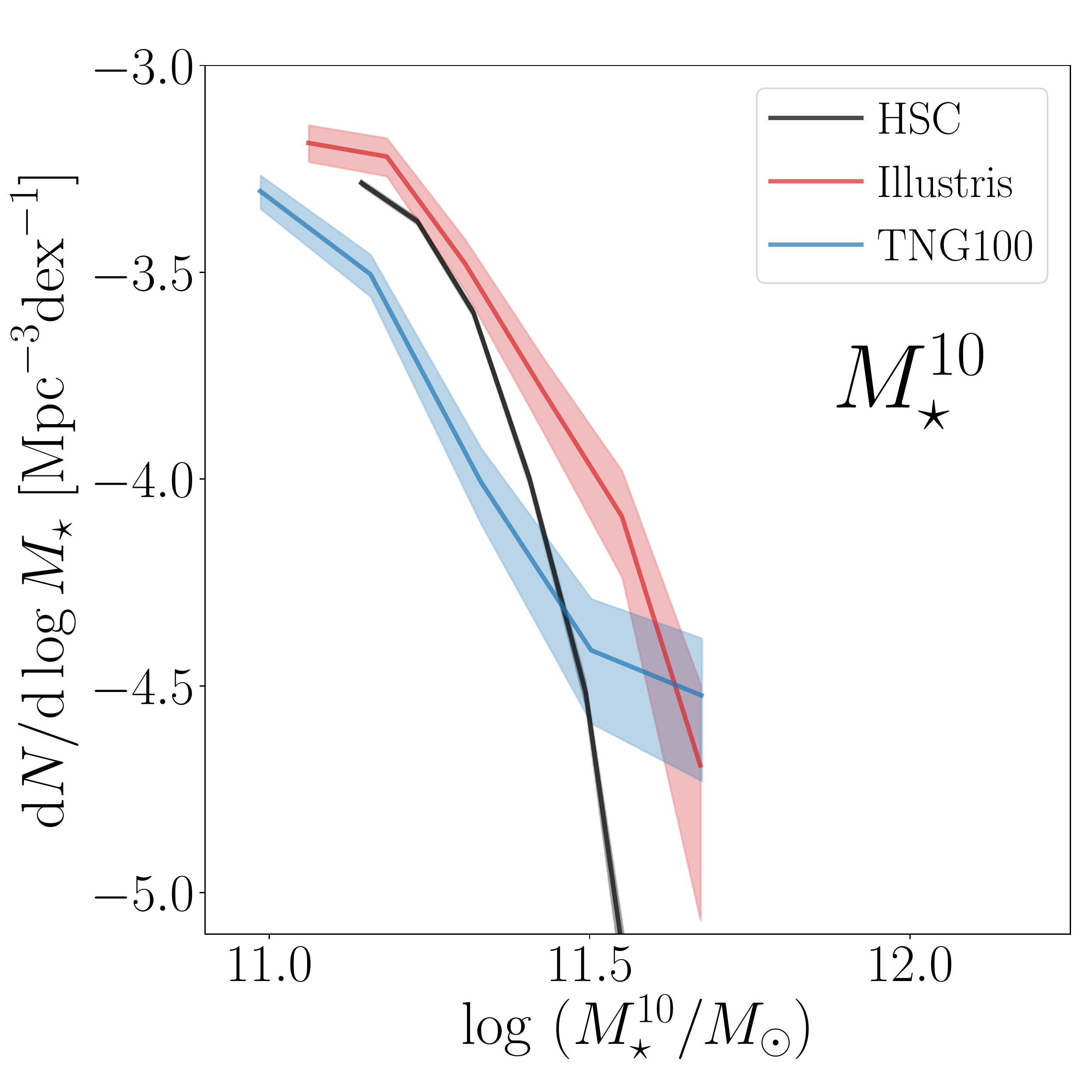}    \includegraphics[width=0.37\textwidth]{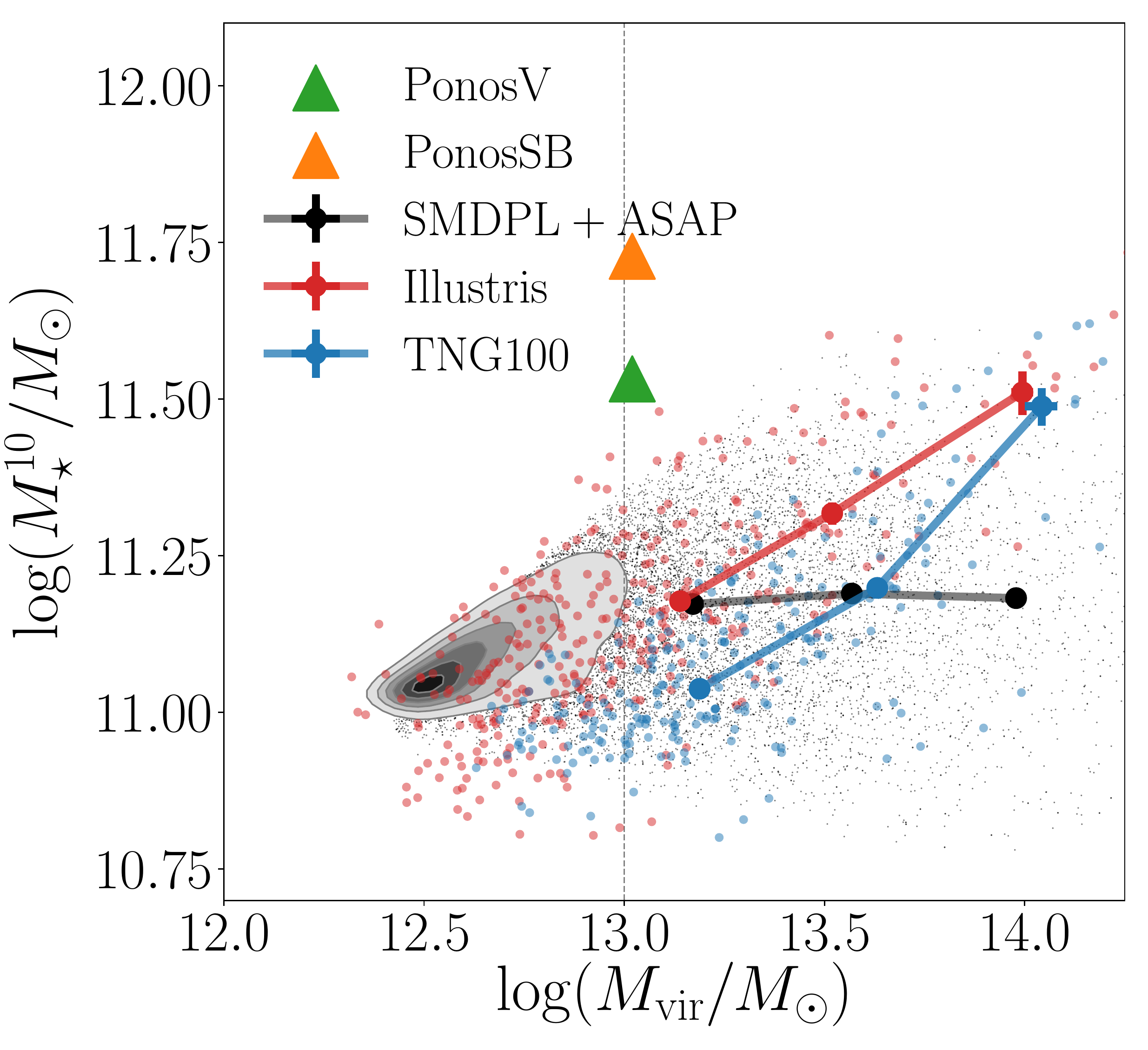} \\
    \includegraphics[width=0.37\textwidth]{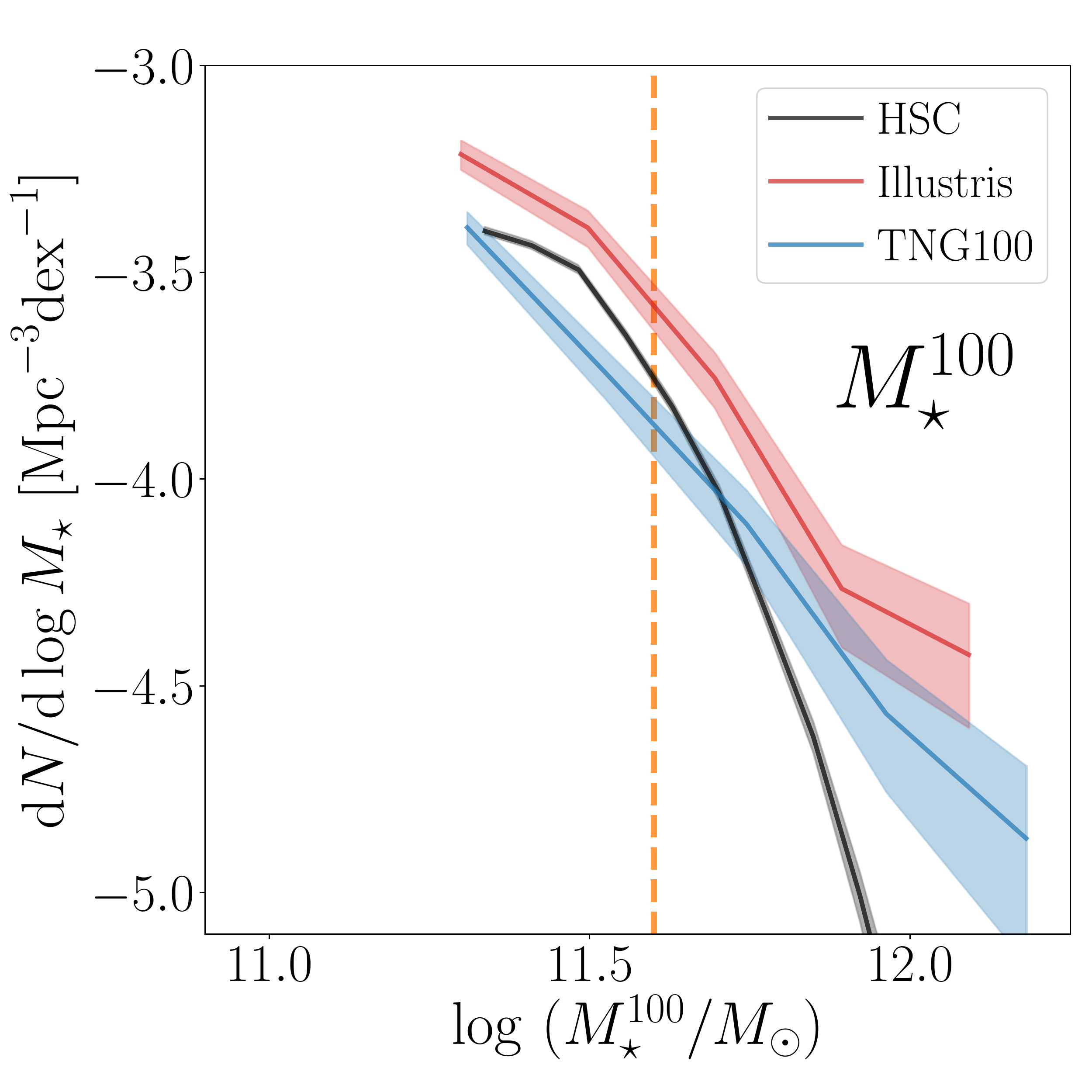}
    \includegraphics[width=0.37\textwidth]{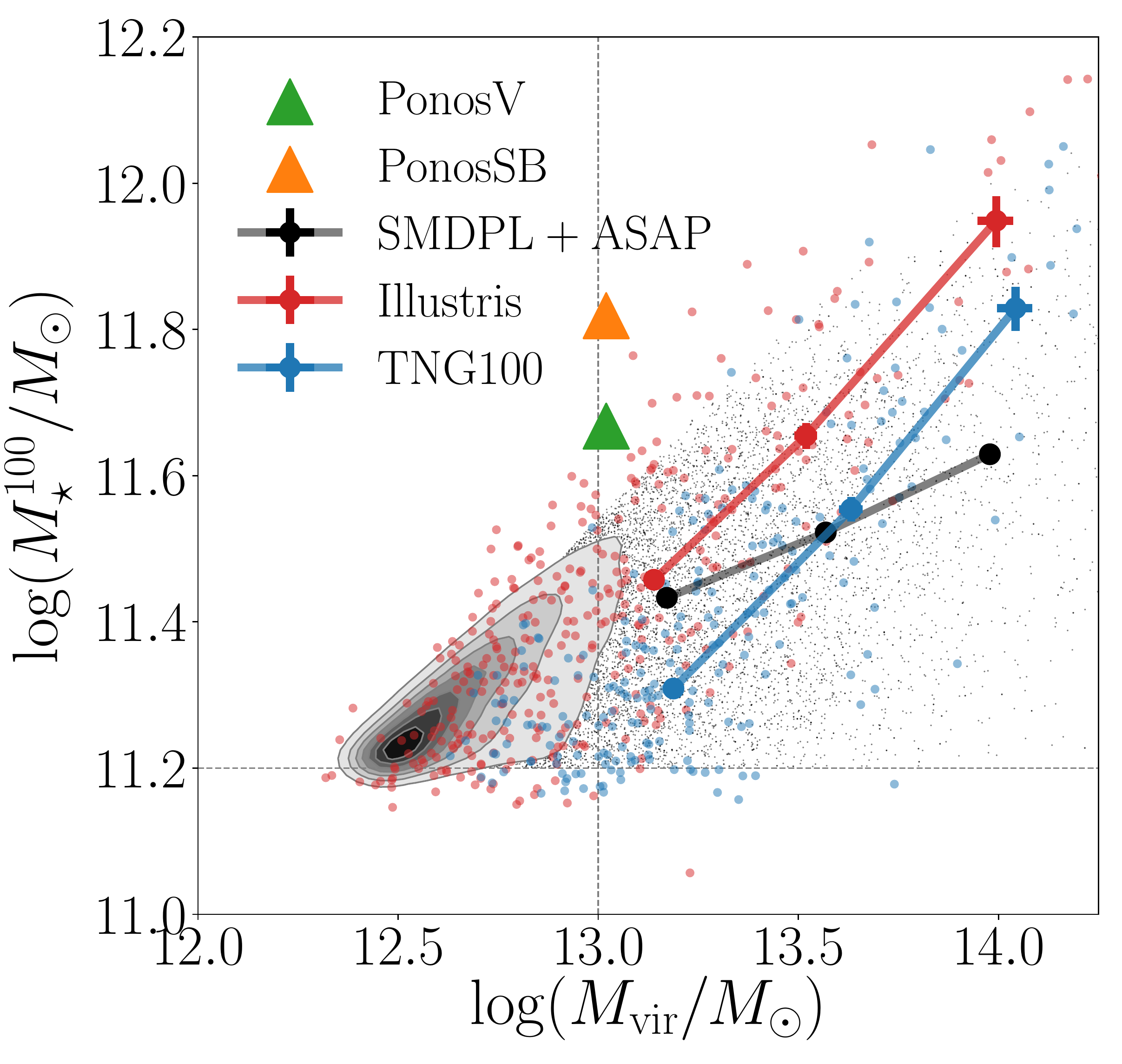} \\
    \includegraphics[width=0.37\textwidth]{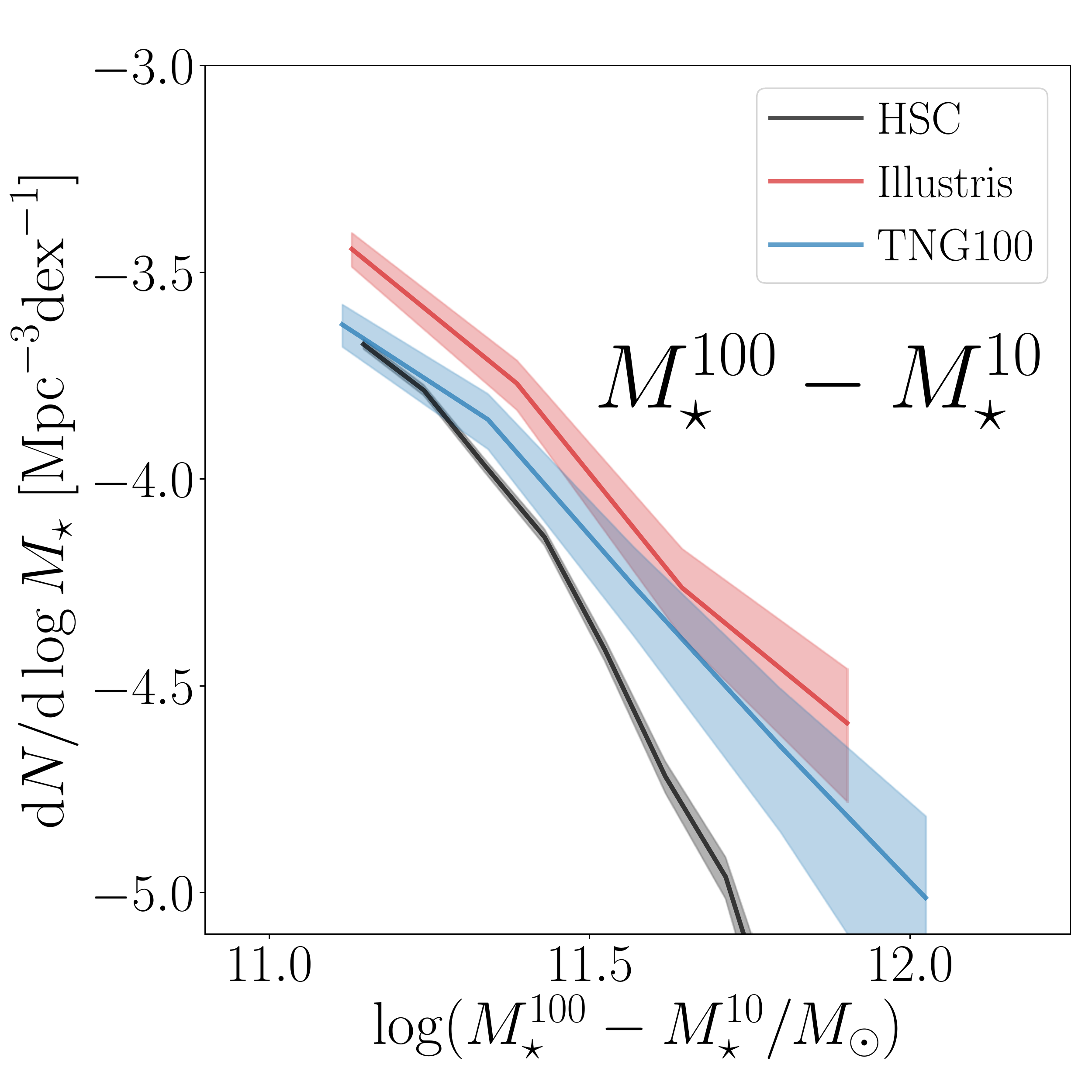}
    \includegraphics[width=0.37\textwidth]{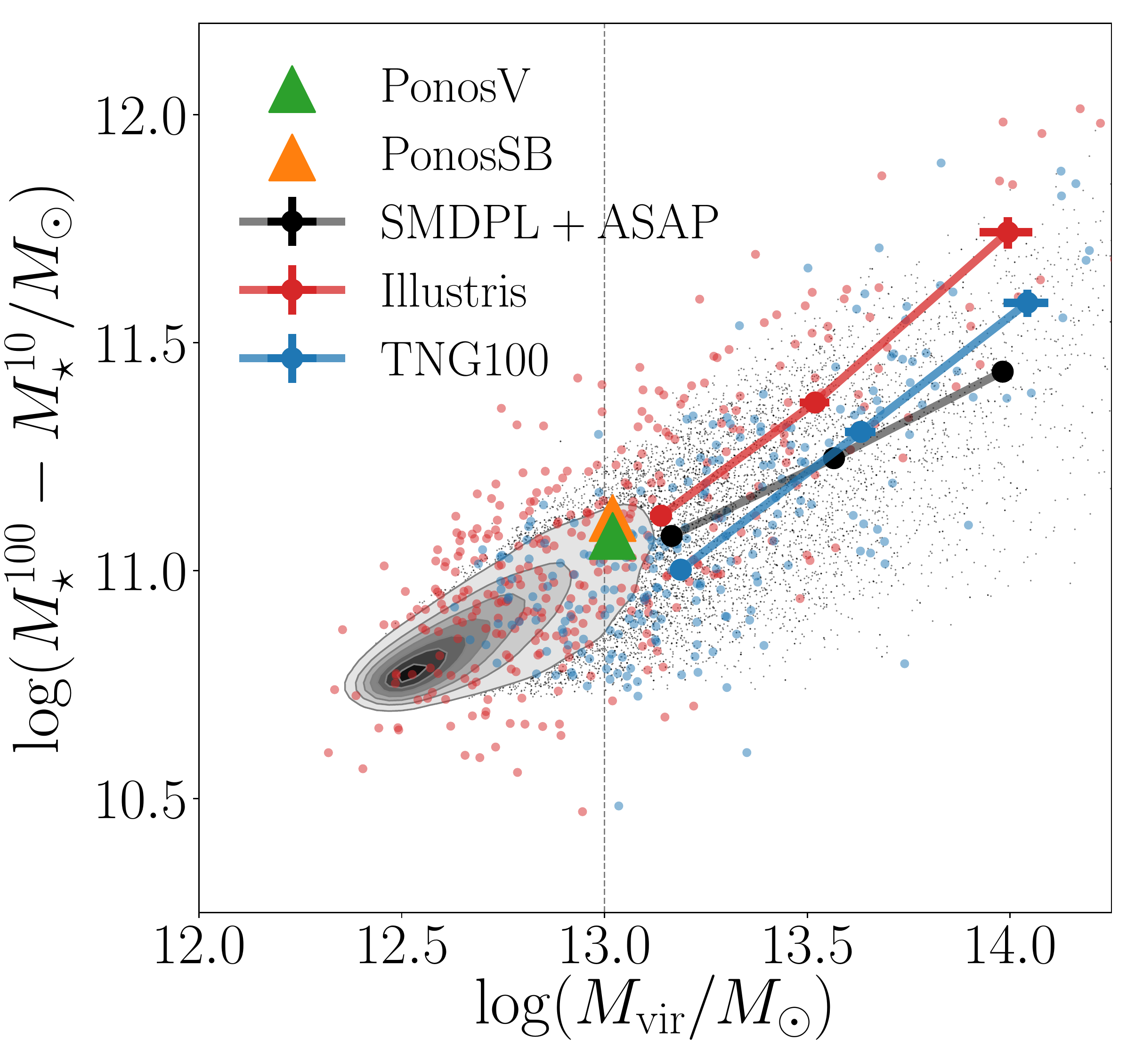}
  
  \caption{Stellar mass functions (left) and stellar vs. halo masses (right), for our $z\sim0.4$ samples using different mass definitions. From top to bottom, the stellar masses shown are  $M_{\star}^{10}$, $M_{\star}^{100}$, and $M_{\star}^{100} - M_{\star}^{10}$ (outer mass). \textbf{Left:} Comparison of SMFs between HSC sample (black), Illustris (red) and TNG100 (blue). Poisson errors are shown in lighter shading for each line. The orange vertical line at $M_{\star}^{100} = 11.6$ corresponds to the completeness limit for our HSC sample. \textbf{Right:} Comparison of stellar vs. halo mass for HSC (black dots and contours), Illustris (red), and TNG100 (blue). PonosV (green) and PonosSB (orange) are also included with triangle markers for comparison. Smaller lighter points are individual galaxies, and larger darker points show median values in bins of halo mass for each sample. Our samples are complete at $\log \left(M_{\mathrm{halo}}\right) > 13.0$ so we only show medians above this limit (grey vertical line). We also show the standard error of the mean as error bars, when they are larger than the marker size. Generally, at lower masses Illustris galaxies are in excess, while TNG100 galaxies are in deficit compared to HSC/SMDPL+ASAP. At higher masses both Illustris and TNG100 show excess stellar mass. Notes: (1) the y-axis scales are not the same in all plots; (2) in the bottom right plot, the Ponos markers have been offset slightly from each other in order for both to be visible.} 
 
  \label{fig:SHMR+SMFs}
\end{figure*}

With a consistent methodology for measuring surface mass density profiles for massive central galaxies in both HSC observations and hydrodynamic simulations, we can now begin to compare and quantify similarities and differences between the data and the simulations. There are several ways to make these comparisons. We begin by investigating differences in the stellar mass function (SMF; \ref{sec:SMF}) and the stellar-to-halo mass relation (SHMR; \S \ref{sec:SHMR}). We also compare the stellar mass density profiles of galaxies in HSC and simulations matched by their ``total" stellar mass ($M_{\star}^{\mathrm{100}}$) in \S \ref{sec:match_by_m_star}, and matched by their halo mass ($M_{\mathrm{\rm vir}}$) in \S \ref{sec:match_by_m_halo}. Finally, to demonstrate that we are indeed comparing at fixed halo mass, we compare weak-lensing profiles in bins of halo mass in \S \ref{sec:weak_lensing_comparisons}.

\subsection{Stellar Mass Functions and Stellar-to-Halo Mass Relation} 
In Figure \ref{fig:SHMR+SMFs} we present both the $z\sim 0.4$ SMF (left column) and the stellar-to-halo mass relation (right column) for our HSC sample compared to Illustris and TNG100, split by stellar mass definitions. The top two panels use $M_{\star}^{\mathrm{r}}$ at $r= \ $10 kpc and 100 kpc for stellar mass. The bottom panel uses the stellar mass between 10 and 100 kpc (representing the outer mass of the galaxy). 

\subsubsection{SMF}\label{sec:SMF}

We find that Illustris galaxies are overly massive at all radial scales and mass bins, particularly at higher masses where there can be an offset of $\gtrsim 0.3$ dex. This discrepancy \citep[see also][]{Genel+2014:Illustris} is likely related to the AGN feedback model, which in the Illustris implementation is now known to not be efficient enough to sufficiently quench star formation \citep{Weinberger+2017, Donnari+2019}. In TNG100, where the AGN feedback model was improved, we see that the SMF more closely matches that of HSC. While the amplitude of the TNG100 SMF compared to Illustris is reduced and results in a better match to observations, the shape of the SMF is nearly the same, much shallower than the observed SMF. In TNG100 this means too few lower mass galaxies and too many higher mass galaxies at all aperture masses (the HSC and TNG100 SMFs cross at $M_{\star}^{\mathrm{10}}\sim10^{11.5}$, $M_{\star}^{\mathrm{100}}\sim10^{11.7}$, and $M_{\star}^{\mathrm{100}}-M_{\star}^{\mathrm{10}}\sim10^{11.2}$). The bottom panel, which uses the outer galaxy mass ($M_{\star}^{\mathrm{100}}-M_{\star}^{\mathrm{10}}$) SMF, shows that both the Illustris and TNG100 galaxies are slightly offset from HSC across all masses, indicating that there is more outer mass in Illustris/TNG100 galaxies than in HSC galaxies.

Figure \ref{fig:SHMR+SMFs} also highlights the fact that the level of agreement between simulations and observations depends on the adopted mass definition (inner mass, total mass, or outer mass). It expands upon the analysis of the galaxy stellar mass functions of Illustris and IllustrisTNG presented in \citealt{Pillepich+2018b:TNG}, where 3D based stellar mass definitions (within 10, 30, 100 kpc and twice the stellar half-mass radius) are contrasted at $z=0$. Here, we further this work by contrasting with high-quality HSC data and by adopting 2D masses which are more directly comparable to data such as from HSC. 

The volume of our HSC sample between $0.3 < z < 0.5$ is approximately $420^3$ Mpc$^3$, compared to $\sim110^3$ Mpc$^3$ for both Illustris and TNG100. For this reason, the errors on the HSC SMF are smaller than for Illustris and TNG100.

\subsubsection{SHMR}\label{sec:SHMR}

Figure \ref{fig:SHMR+SMFs} compares the SHMR between the observed and simulated galaxies. Rather than use HSC mass measurements directly, we use the values of the \asap\ model, applied to the SMDPL simulation. This corresponds to the best fit model to the HSC data (HSC SMF, $M_{\star}^{10}$, $M_{\star}^{100}$, and weak lensing). This means that we use the full distribution of halo masses from SMDPL, $P(M_{\rm vir}|\overline{M}_{\rm vir})$, and stellar masses are derived from those halo masses using \asap. By using this best fit model, we can display not only the mean SHMR but also the scatter in the SHMR that best fits the HSC data. We have checked that using a top hat bin in $M_{\mathrm{vir}}$ versus using the full distribution $P(M_{\rm vir}|\overline{M}_{\rm vir})$ does not impact these results. 

Figure \ref{fig:SHMR+SMFs} shows that Illustris galaxies in halos with $\log(\mathrm{M}_{\rm vir})\gtrsim 13.5$ contain too much stellar mass for all aperture masses. For Illustris galaxies in halos of $\log(\mathrm{M}_{\rm vir})\sim 13.5$ ($\log \mathrm{M}_{\rm vir} \sim 14$), there is a constant offset of  $\sim0.1$ dex ($\sim0.3$ dex), corresponding to a $\sim 30\%$ ($\sim 100\%$) difference in stellar mass. Galaxies in less massive Illustris halos ($\log \mathrm{M}_{\rm vir} \sim 13$) tend to match SMDPL+\asap\ well, and are offset by less than 0.04 dex ($< 10\%$).

TNG100 galaxies show a slightly different result, but similarly consistent across aperture mass definitions. In general, the lowest mass halos host galaxies with less stellar mass than SMDPL+\asap, the highest mass halos have more stellar mass than SMDPL+\asap, and galaxies in the middle halo mass bin have stellar masses consistent with SMDPL+\asap\ (offset by less than 0.05 dex).  For galaxies in halos of $\log\left(\mathrm{M}_{\rm vir}\right)\sim 13$ there is a constant deficit of $\sim0.1$ dex in stellar mass, corresponding to a $\sim 25\%$ difference compared to SMDPL+\asap. In the most massive halos ($\log \mathrm{M}_{\rm vir} \sim 14$), there is an excess of $\sim0.15-0.30$ dex which corresponds to a $\sim 40-90\%$ difference from SMDPL+\asap, this disagreement being larger in the inner regions of galaxies. In fact, TNG100 returns an outer stellar mass ($M_{\star}^{100}-M_{\star}^{10}$) that is consistent with the SMDPL+\asap~ results to better than 0.15 dex across the halo mass range studied here. The fact that Illustris and TNG100 show significant offsets in some halo mass bins in either the inner ($M_{\star}^{10}$) or outer ($M_{\star}^{100}-M_{\star}^{10}$) stellar mass, or both, suggests that the offsets may be associated to both in-situ and ex-situ stellar mass processes. We will discuss these implications in \S \ref{sec:discussion}.

If $\alpha$ is the slope of the high mass end of the SHMR with $M_{\star}\propto M_{\rm vir}^{\alpha}$ then TNG100 appears to have a steeper value of $\alpha$ compared to SMDPL+\asap, with differences most notable in the inner regions of galaxies. This could suggest that while the overall strength of AGN feedback better matches observations in TNG100, the halo mass dependence may still need adjusting. Accretion of low mass satellites also play a role at these mass scales (see discussion in \S \ref{sec:dual_impact}).

\begin{figure}
    \centering
    \includegraphics[width=0.49\textwidth]{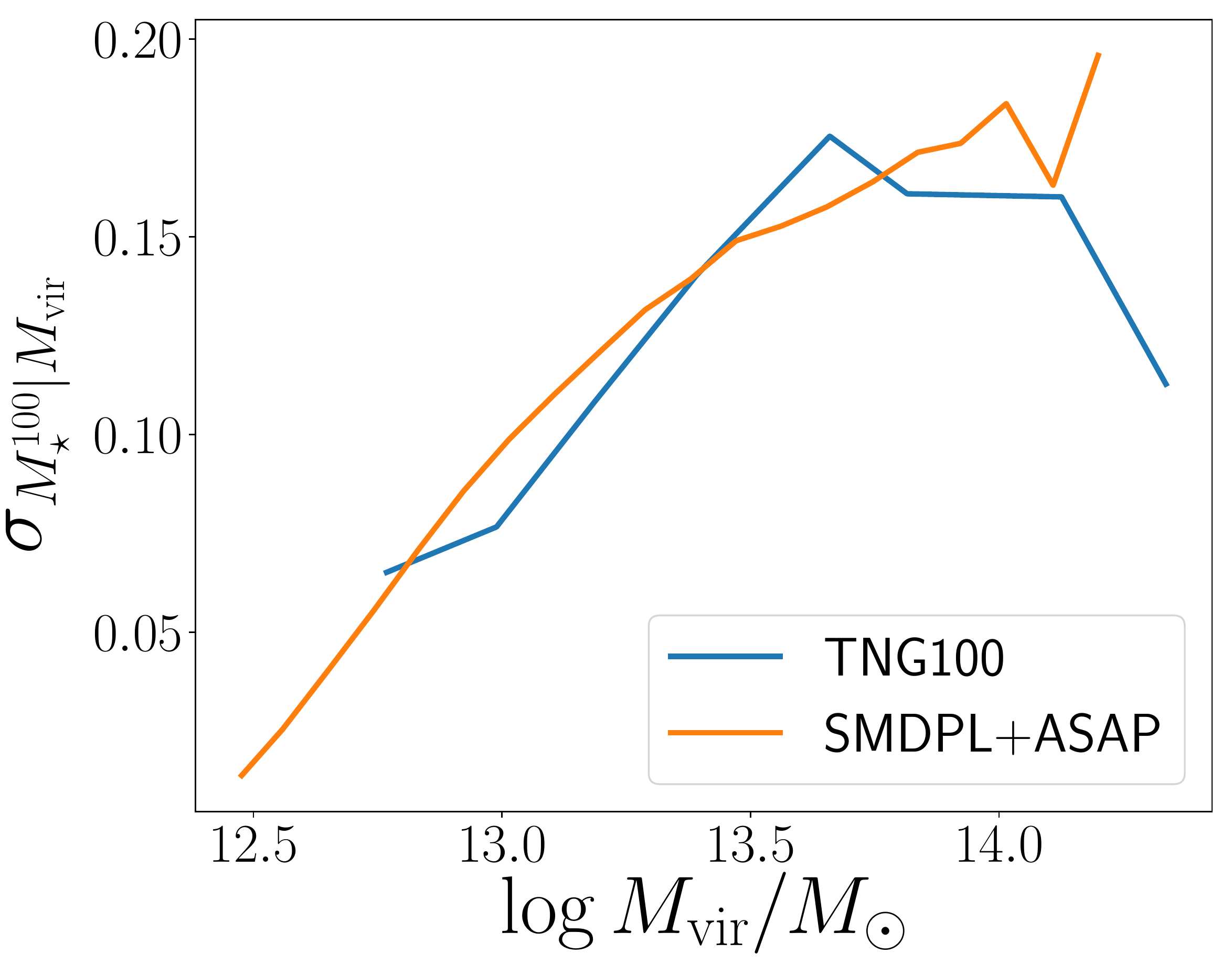}
    \caption{Comparison of the scatter in stellar mass at fixed halo mass for galaxies in TNG100 compared to SMDPL+\asap. In the halo mass range which we investigate in this work ($13.0<\log M_{\mathrm{vir}}<14.25$) the scatter in stellar mass is comparable between the two samples. This suggests that the scatter is not driving the difference in the SHMR.
    }
    \label{fig:scatter}
\end{figure}

We have also checked that the scatter in stellar mass at fixed halo mass (Figure \ref{fig:scatter}) is comparable for TNG100 and SMDPL+\asap\ for most of the halo mass range which we investigate in this work ($13.0<\log M_{\mathrm{vir}}<14.25$). This suggests that the dominant driver in the difference we see in the SHMR (Figure \ref{fig:SHMR+SMFs}) is the change in slope rather than the scatter.

Both Ponos galaxies have inner stellar masses that are higher than any galaxy in Illustris, TNG100, or SMDPL+\asap~ at similar halo mass (a difference of $> 0.25$ dex from the median SMDPL+\asap~ value). The outer mass, on the other hand, appears to be comparable to that of the other samples (a difference of $\lesssim 0.1$ dex from the median SMDPL+\asap~ value). The significance of this will be discussed further in \S \ref{sec:dual_impact}.

\subsection{Density Profile Comparison at Fixed Aperture Stellar Mass} \label{sec:match_by_m_star}

\begin{figure*}
\begin{center}
\includegraphics[width=18cm]{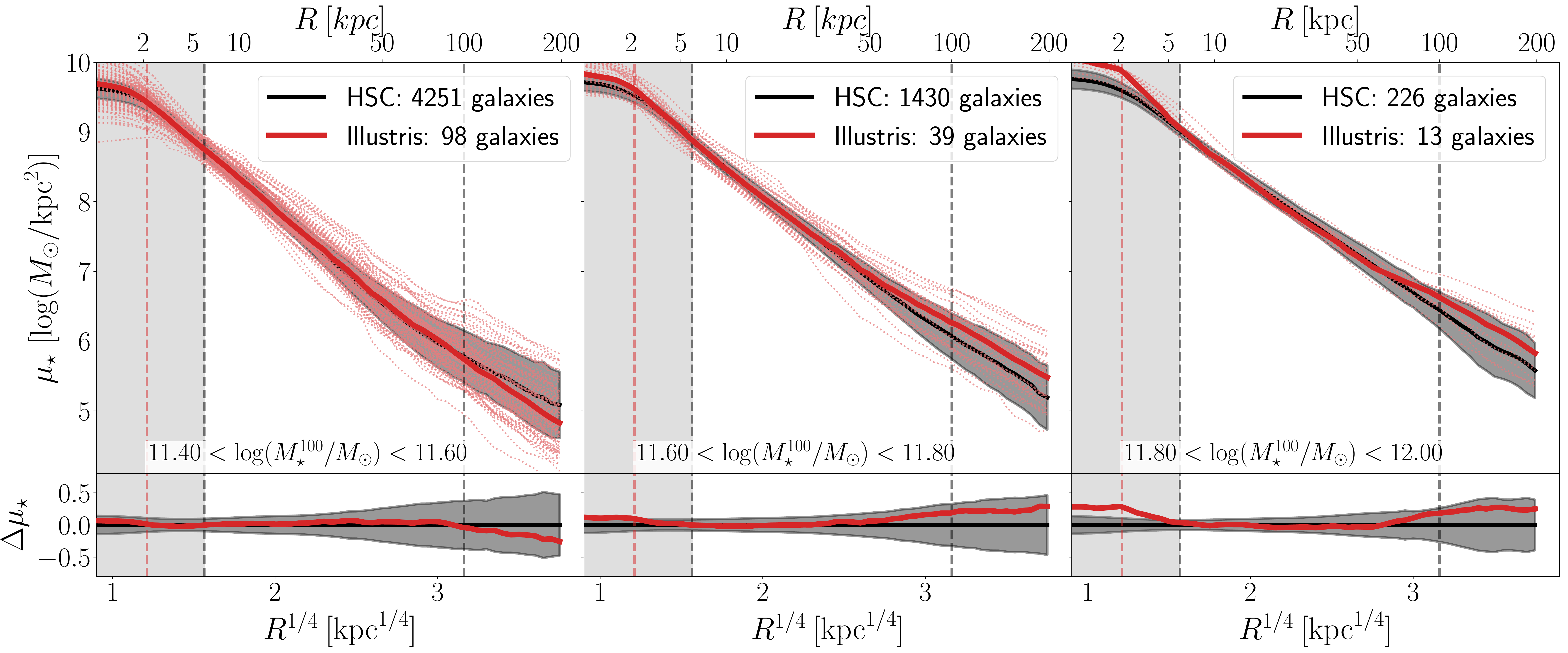}
\includegraphics[width=18cm]{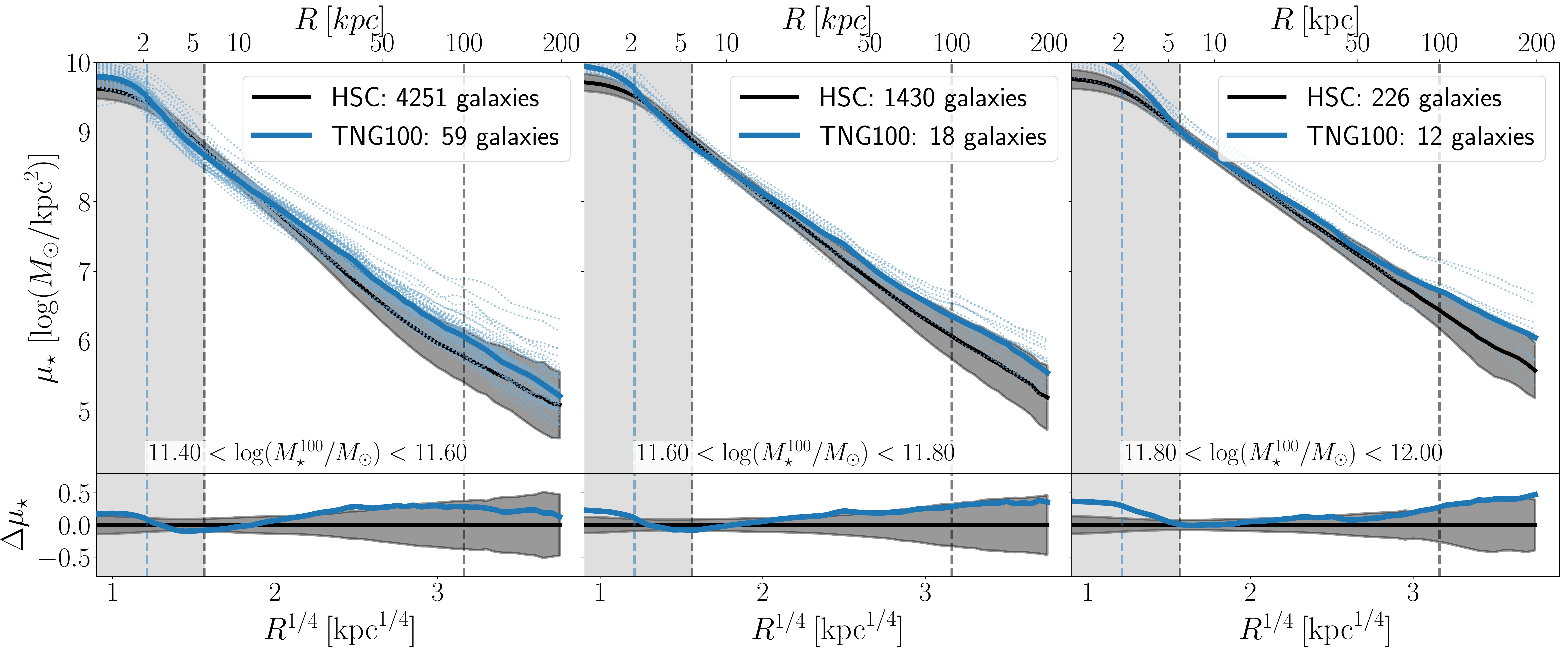}
\caption{Stellar mass density profiles of $z\sim0.4$ galaxies in HSC (black), Illustris (top in red), and TNG100 (bottom in blue) matched by $M_{\star}^{\mathrm{100}}$. In all panels, the lighter dotted lines are individual galaxies in the simulations and the thicker lines are medians. Black solid lines are the median HSC stellar mass density profiles, with the 1$\sigma$ widths of the distribution in grey. The shaded grey rectangular area between 0-6 kpc represents the region in which the HSC profile is sensitive to instrument PSF effects. The vertical line at 100 kpc indicates the maximum extent to which we are confident in the background subtraction of our HSC profiles \citep[see][]{Huang+2018b}. The colored vertical lines at $\sim 2.1$ kpc show 3x the force resolution of the stellar particles in Illustris and TNG. The bottom panel of each figure also shows the residual difference between the median HSC profile and the median simulation profile. Illustris and TNG100 have slightly shallower profile slopes compared to HSC.}
\label{fig:matched_by_m100}
\end{center}
\end{figure*}

In this section, we compare the galaxy samples in bins of stellar mass ($M_{\star}^{\mathrm{100}}$). Figure \ref{fig:matched_by_m100} shows the stellar mass density profiles of HSC galaxies and of simulated galaxies (Illustris in red and TNG100 in blue) in three bins of $M_{\star}^{\mathrm{100}}$. The lighter thinner dotted lines are the profiles of individual simulated galaxies, while the thicker lines are the median profiles. Because of the large number of HSC galaxies in our sample, we only show the median profile (in black) and the 1$\sigma$ width in gray. When matching by stellar mass, the overall amplitudes will match by design (since by construction integrated mass within 100 kpc is identical). This comparison therefore serves mainly the purpose of understanding whether or not the profiles have similar slopes. We also note that the completeness of our HSC sample is only $\sim65\%$ in the lowest mass bin ($10^{11.4}<M_{\star}^{\mathrm{100}}<10^{11.6}$), but close to $100\%$ at higher masses.

For both Illustris and TNG, in every mass bin, the median profiles of the simulated galaxies lie at or within the 1$\sigma$ width of the distribution of HSC galaxy profiles, suggesting that the slopes of the profiles are in relatively good agreement. Our finding supports recent work on the density profiles of early-type galaxies in TNG100, which show that the power-law density slope of these galaxies is largely in agreement with different observations in different bins of stellar mass \citep{Wang+2018}. 

The offset of the median Illustris stellar mass from the median HSC mass in bins of stellar mass, is small ($\sim 0.01$ dex) in all mass bins and at all mass aperture radii. In TNG100, the mass offsets are of similar magnitude as Illustris in the inner regions of the galaxies ($r< 10$ kpc), but become larger in the outer regions ($r> 10$ kpc). Overall, Illustris and TNG100 tend to have slightly shallower profile slopes than HSC, with differences increasing towards the high mass end.

\subsection{Density Profile Comparison at Fixed Halo Mass} \label{sec:match_by_m_halo}

\subsubsection{Illustris and TNG100}

\begin{figure*}
\begin{center}
\includegraphics[width=18cm]{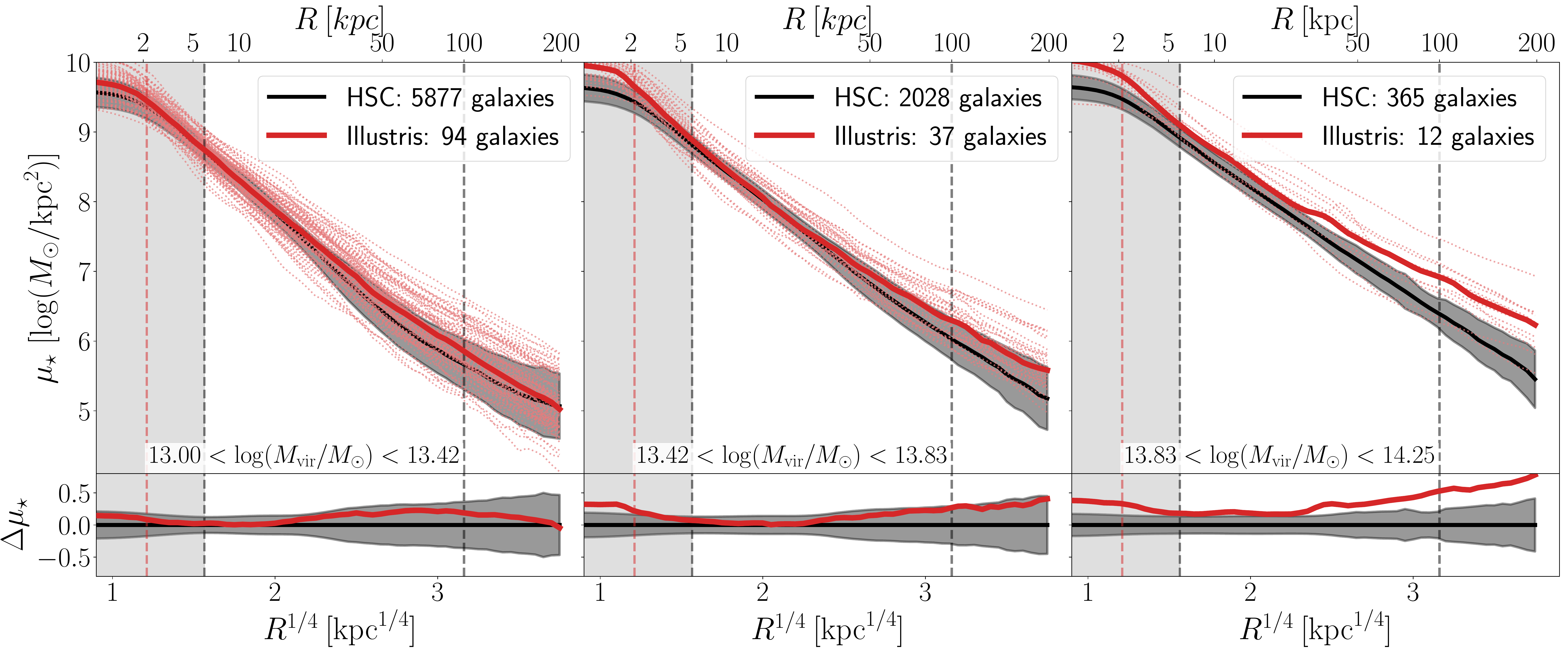}
\includegraphics[width=18cm]{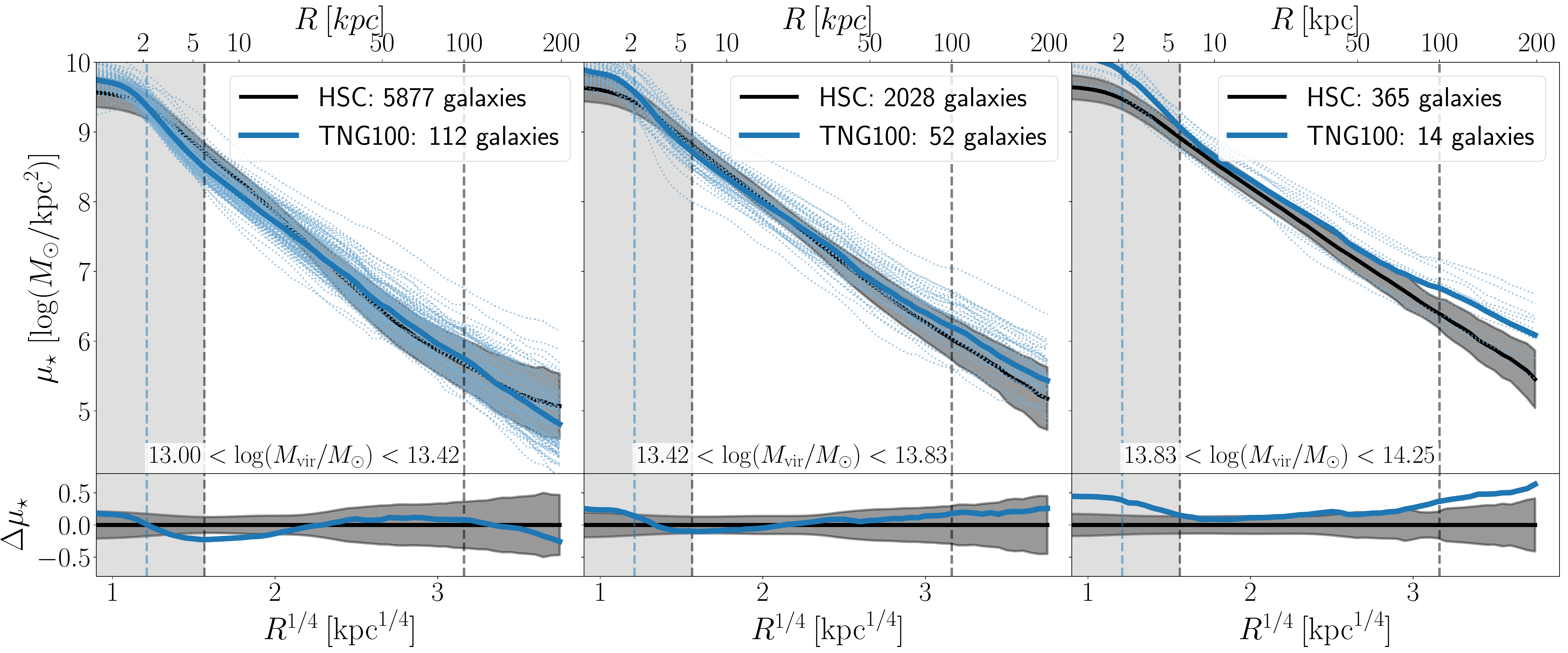}
\caption{Same as Figure \ref{fig:matched_by_m100}, but in bins of $M_{\mathrm{\rm vir}}$. When matched by halo mass, the stellar mass density profiles of Illustris/TNG100 galaxies are relatively well matched to HSC. There is an excess of stellar mass in the outer regions which increases with halo mass. There is also a noted deficit of stellar mass in the inner regions of TNG100 galaxies in lower halo masses.}
\label{fig:matched_by_halo_mass}
\end{center}
\end{figure*}

A more informative way to compare galaxies in HSC with those from simulations is to match them by dark matter halo mass. This is a key and unique novelty in this paper that is enabled by the weak lensing measurements from HSC. Figure \ref{fig:matched_by_halo_mass} shows the median stellar mass density profiles of HSC compared to Illustris (red lines in top panels) and TNG100 (blue lines in bottom panels) in bins of halo mass (same format as Figure \ref{fig:matched_by_m100}). We select three equally-spaced bins in the mass range for which we expect to be complete in our HSC sample ($M_{\mathrm{\rm vir}} > 10^{13} M_{\odot}$). For HSC we use $\overline{M}_{\rm vir}$ as described in \S \ref{sec:measurements:data:weak-lensing}. For Illustris/TNG100 galaxies, we use $M_{\mathrm{vir}}$ (labeled \texttt{Group\_M\_TopHat200}) as measured directly by the halo finder from the simulation output.

Figure \ref{fig:matched_by_halo_mass} shows that in general, both Illustris and TNG100 galaxy profiles agree relatively well with HSC galaxies. For both Illustris and TNG100, at larger halo masses there are larger offsets from HSC. In halos of $M_{\star} \sim 10^{14} M_{\odot}$, the stellar mass profiles of TNG100 (Illustris) are offset from HSC by about 0.3 ($\sim0.5$) dex in the very outskirts ($\sim100$ kpc). In the same high mass bin, a similar significant offset in stellar mass is manifest also in the innermost regions ($< 6$ kpc) for both Illustris and TNG100, while TNG100 exhibits stellar mass profiles in agreement with observed galaxies to better than 0.1 dex across the 10-80kpc range. In the lower halo mass bins ($M_{\star} \sim 10^{13} M_{\odot}$ ), Illustris galaxies have an excess of up to $\sim0.1$ dex in both the inner and outer regions. On the other hand, lower halo mass TNG100 galaxies display a notable deficit ($\sim 0.1$ dex) in stellar mass in the inner regions (around 6-10 kpc galactocentric distances). Overall, these differences translate into the result that Illustris/TNG100 display steeper SHMRs than HSC as is also seen in Figure \ref{fig:SHMR+SMFs}, but with a remarkable agreement in the stellar mass distributions between HSC and TNG100 galaxies in $M_{\star} \sim 10^{13.5} M_{\odot}$ haloes.

\subsubsection{Ponos Simulation} \label{sec:ponos_results}

Our two Ponos galaxies are different hydrodynamic realizations of the same mass halo ($M_{\mathrm{\rm vir}} = 1.04 \times 10^{13} M_{\odot}$). In this paper, they serve as an interesting counterpart to Illustris/TNG100 because they do not include AGN feedback and so help to inform our understanding of the impact of AGN feedback on galaxy mass profiles. 

We compare Ponos galaxies with HSC at fixed halo mass. We select 75 HSC galaxies from our full sample such that $1.03 \times 10^{13} M_{\odot}< \overline M_{\mathrm{\rm vir}}< 1.05 \times 10^{13} M_{\odot}$, where $\overline M_{\mathrm{\rm vir}}$ is calculated using the mean $M_{\rm vir}-M_{\star}$ relation as given by the ASAP model presented in \citet{Huang+2019} (see \S \ref{sec:measurements:data:weak-lensing}). It should be noted that the median mass of our HSC comparison subsample ($\log M_{*}^{\mathrm{100}}/M_{\odot}=11.35$) is below our completeness limit and so the comparison in this regard is not as accurate as in previous sections. 

\begin{figure*}
      \centering 
      \includegraphics[width=0.475\textwidth]{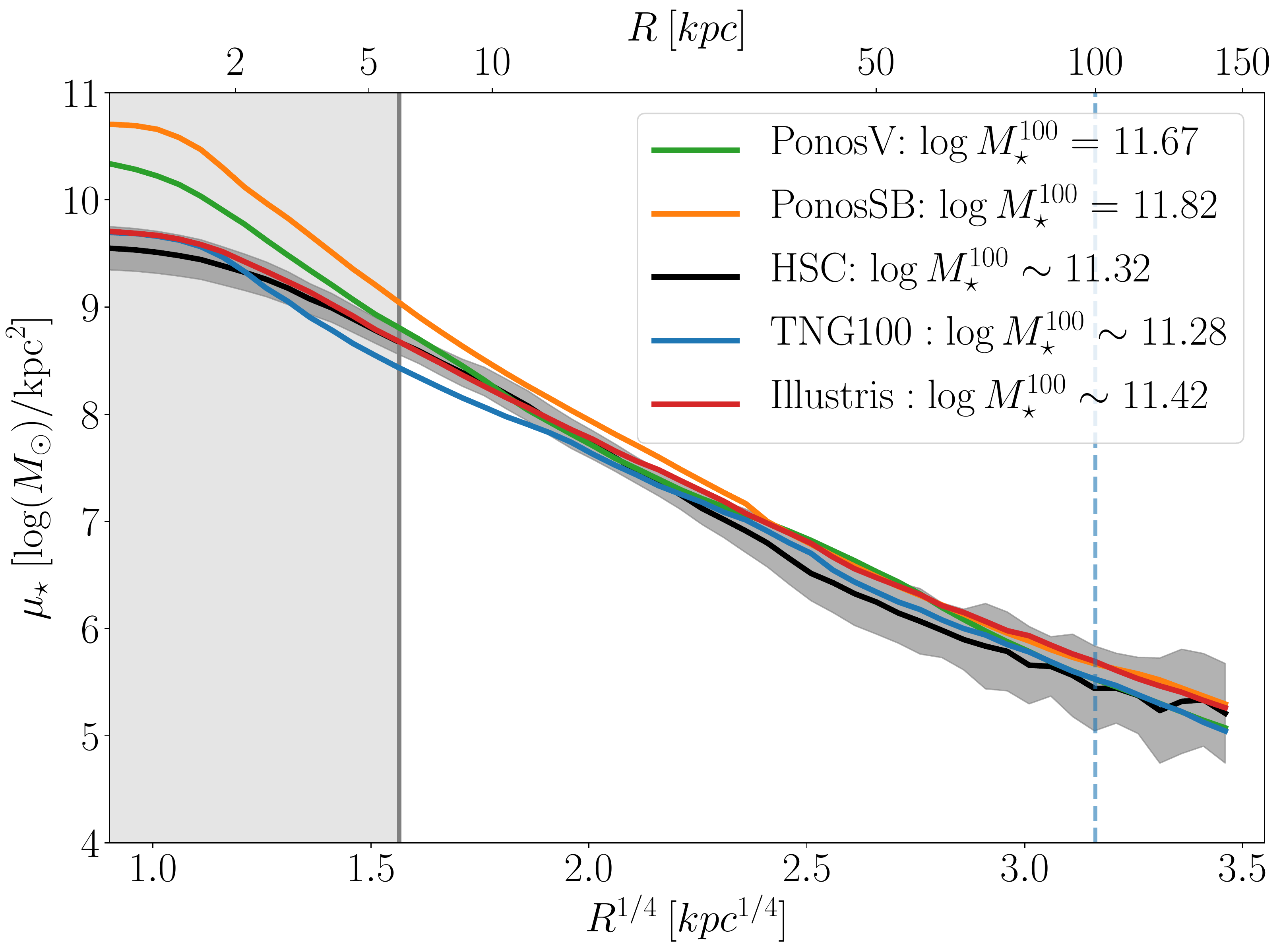}
      \includegraphics[width=0.49\textwidth]{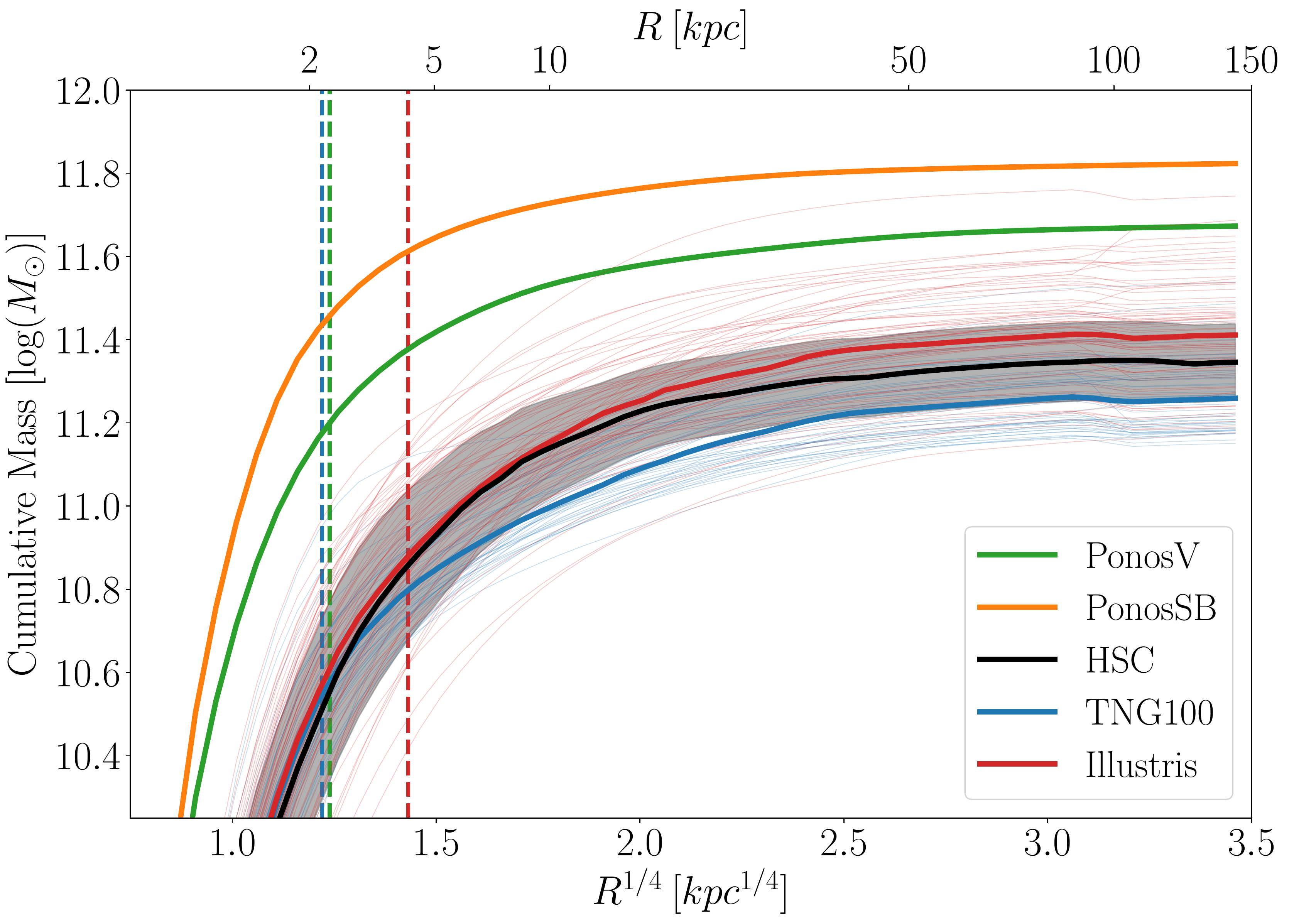}
      \caption{\textbf{Left:} Surface stellar mass density profiles of our two Ponos galaxies (green and orange), and our comparison subsamples of Illustris (red), TNG100 (blue), and HSC (gray) galaxies at matched halo mass ($M_{\rm vir}\sim 10^{13}M_{\odot})$. The darker black line is the median profile of HSC galaxies. The Ponos simulations do not include AGN feedback which results in over-cooling and a large excess of mass compared to HSC at $r\lesssim20$ kpc. \textbf{Right: } Comparison of the cumulative stellar mass profiles of the same galaxies as in the left panel. The lack of AGN feeback in Ponos results in very dense inner regions which contain most of the stellar mass of the galaxy, seen here with a very steep inner slope at $r\lesssim20$ kpc. The outer regions of galaxies in Ponos, TNG, and Illustris have comparable slopes. The vertical lines show three times the DM particle softening length in Ponos (green at 2.4kpc), TNG100 (blue at 2.2 kpc), and Illustris (red at 4.3 kpc). This comparison shows that the stellar profiles of massive galaxies contain information about the physical scale of AGN feedback -- this work suggests the physical scale of AGN feedback in massive galaxies to be of order $r\lesssim20$ kpc.}
      \label{fig:ponos}
  \end{figure*}

Figure \ref{fig:ponos} shows a comparison of the stellar mass surface density profiles for HSC, Ponos, Illustris, and TNG100. Even though we only have 2 Ponos galaxies it is clear from Figures \ref{fig:SHMR+SMFs} and \ref{fig:ponos} that Ponos galaxies are offset compared to HSC/Illustris/TNG100. The very inner regions of Ponos galaxies display an extreme excess of mass compared to HSC, likely resulting from the lack of AGN feedback in these simulations. This impacts total stellar mass and results in a $\sim0.3-0.5$ dex excess in $M_{*}^{\mathrm{100}}$ compared to HSC. Ponos galaxies suffer from the traditional over-cooling problem and this is primarily manifested in the inner region of their profiles. Other analyses of simulated galaxies without AGN feedback show a similar excess of stellar mass for a given halo mass \citep[e.g.][]{Martizzi+2012a, Martizzi+2014, Vogelsberger+2013, Pillepich+2018a}. The stellar mass excess seen in Ponos is on scales with $r\lesssim$ 20 kpc but scales larger than $r\gtrsim$ 20 kpc are in good agreement with HSC/Illustris/TNG100. 

\subsection{Comparison of Weak-Lensing Profiles} \label{sec:weak_lensing_comparisons}

The main results in this paper stem from our ability to perform comparisons at fixed halo mass. In order to demonstrate the validity of our halo mass matching procedure, we also show direct comparisons of the weak lensing signals in HSC and TNG100.

\begin{figure*}
\begin{center}
\includegraphics[width=18cm]{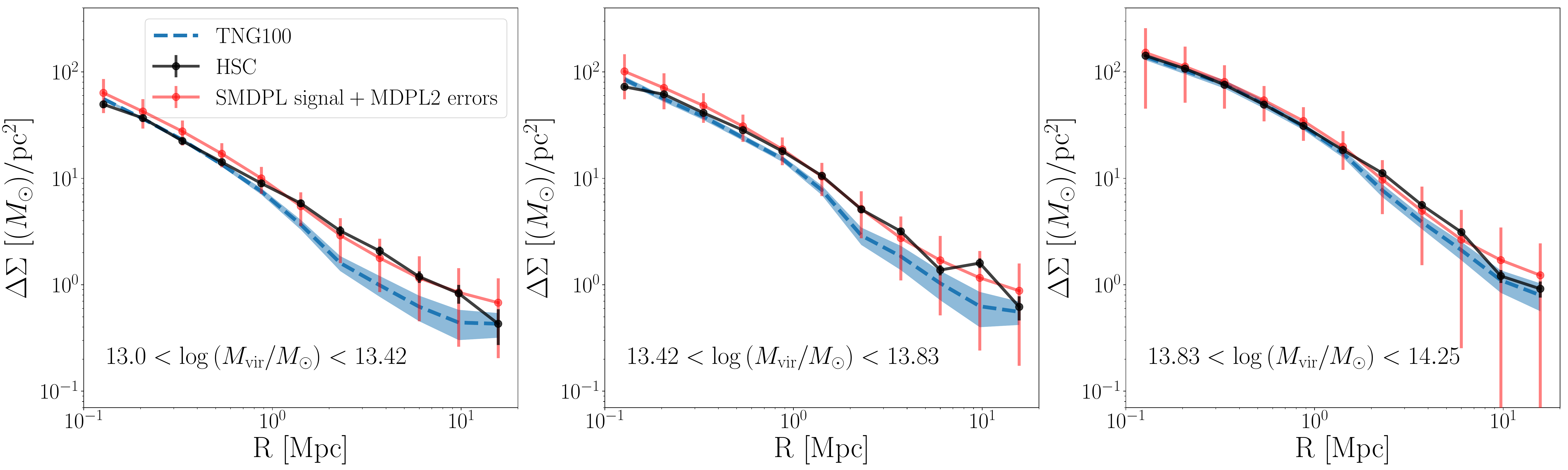}
\caption{Comparison between the galaxy-galaxy weak lensing $\Delta\Sigma$ profiles for HSC and TNG100 in bins of halo mass. Galaxies in HSC are binned by $\overline{M}_{\rm vir}$. The red line shows the best-fit \texttt{ASAP} model in the SMDPL simulation (fits the HSC data by design as described in \citealt{Huang+2019}). For TNG100, we create a sample that is matched in halo masses by drawing from the full probability distribution $P(M_{\rm vir} | \overline{M}_{\rm vir})$. The volume probed by TNG is smaller than in HSC. Red error bars show the sample variance on $\Delta\Sigma$ for TNG100 as calculated using MDPL2 (see \S \ref{sec:weak_lensing_comparisons}). Lighter blue shading is an alternative error estimate based on jackknife resampling of the TNG100 galaxies in each mass bin. In the lowest mass bin in our TNG100 sample there are 82 galaxies, 35 galaxies in the middle bin, and 13 galaxies in the highest mass bin. This figure shows that the weak lensing profiles match well within the errors thus demonstrating that our method for assigning halo mass in HSC is effective.}
\label{fig:weak-lensing}
\end{center}
\end{figure*}

We bin galaxies in HSC according to $\overline{M}_{\rm vir}$ and measure the stacked g-g lensing observable, $\Delta\Sigma$. The results are displayed in Figure  \ref{fig:weak-lensing}. To demonstrate the validity of the \texttt{ASAP} model, we also show the weak lensing signal as measured for the best-fit \texttt{ASAP} applied to the SMDPL simulation (see \S \ref{sec:measurements:data:weak-lensing})

For TNG100, $\Delta\Sigma$ was measured with \texttt{Halotools} \citep[v0.7;][]{halotools} using particle data. To estimate the sample variance error on $\Delta\Sigma$ for TNG which is smaller in volume compared to HSC, we use samples drawn from the MultiDark Planck 2 simulation \citep[MDPL2;][]{Klypin+2016:MDPL2}. We divide the full 1 Gpc/h volume of MDPL2 into several 75 Mpc/h volumes corresponding to the box size of TNG100 and measure the standard deviation of the weak-lensing signal across the various smaller volumes in each halo mass bin. The red error bars shown in Figure \ref{fig:weak-lensing} corresponds to this standard deviation at each radial bin across the entire MDPL2 volume. The HSC weak lensing profile and the one from TNG100 are in good agreement given the sample variance errors associated with TNG100. This suggests that our method for assigning a halo mass to HSC galaxies is effective. More precise comparisons at fixed halo mass will necessitate larger volume hydro simulations.

It is worth noting that in TNG100 we only consider central galaxies, but in observation, we could have $\sim 10\%$ satellite galaxy contamination in our sample, which could affect the inner \ds profile. However, we have checked that the impact of satellites on $\Delta\Sigma$ is negligible compared to the TNG errors on $\Delta\Sigma$ (Huang et al. in prep).

\section{Discussion}\label{sec:discussion}

We have shown that the full stellar mass surface density profiles of massive galaxies are more informative and general than summarizing galaxies by a single number for ``stellar mass". We have compared new data from HSC with galaxy formation models from three different suites of hydrodynamic simulations and find that while there is impressive agreement with some simulations, there are still important differences to point out. We now discuss some explanations for our results and how they may present a new way to constrain feedback models.

\subsection{The Dual Impact of AGN and Stellar Feedback on the Stellar Mass Profiles of Massive Galaxies} \label{sec:dual_impact}
\begin{figure}
    \centering
    \includegraphics[width=\linewidth]{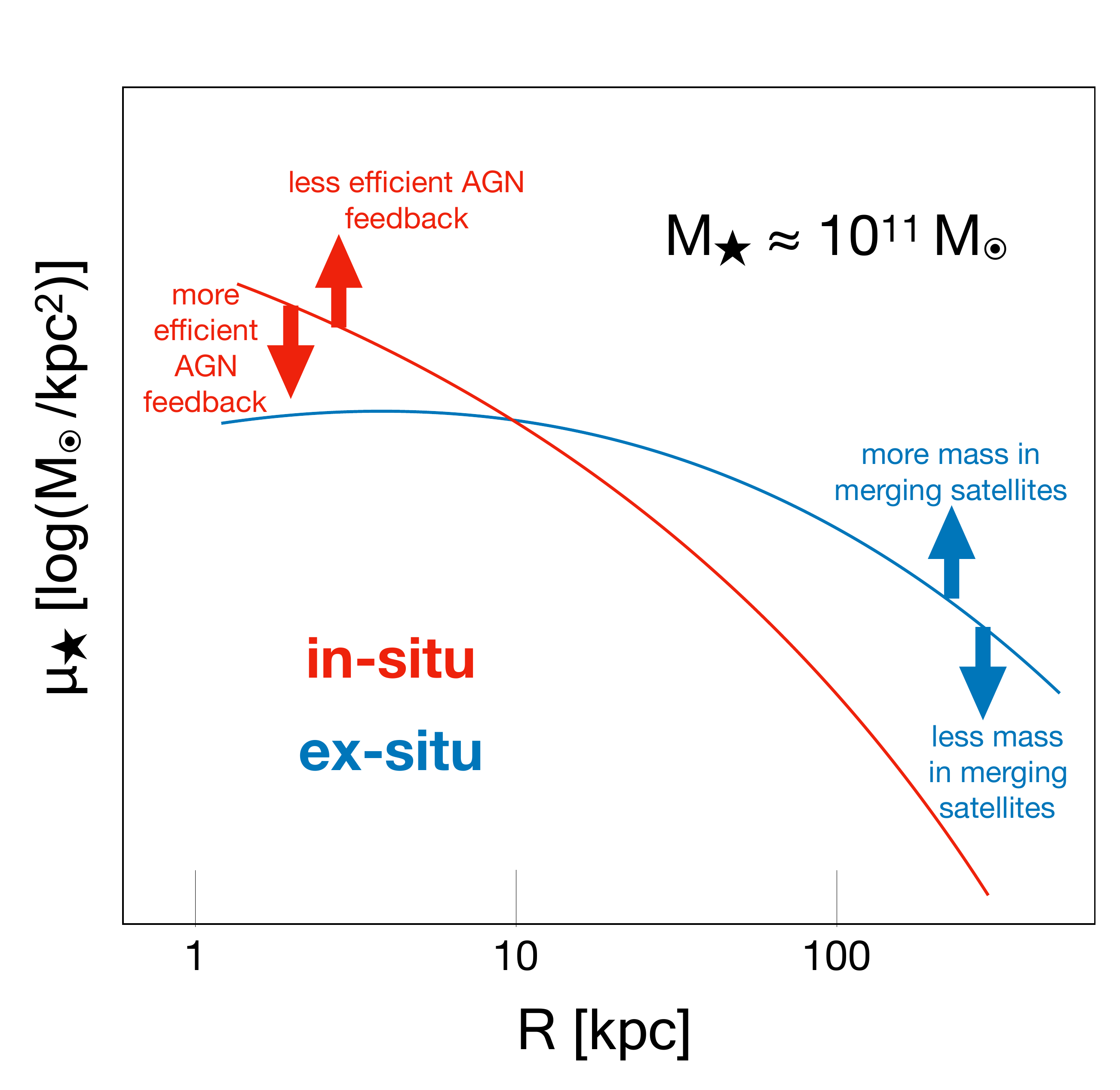}
    \caption{Schematic of the effect of different physical processes on the stellar mass density profile of a galaxy with $M_{\star}\sim 10^{11} M_{\odot}$. The in-situ stellar component is shown in red, while the ex-situ is in blue. The arrows point in the direction that the process in the corresponding color would change the profile. Overcooling in massive halos (due to lack of feedback processes) will result in an increase in mass in the inner regions. Overcooling in small galaxies at early times will result in an increased mass for those galaxies. These small galaxies are eventually accreted to form the ex-situ component resulting in an increase of mass in the outer regions of massive galaxies.}
    \label{fig:toy_profiles}
\end{figure}
One way to explain the mass offsets of galaxies from Illustris/TNG100 is by appealing to the feedback models in the simulations. Feedback regulates the rate of star formation and the distribution of stellar mass within a galaxy. For massive galaxies ($M_{\rm halo} \gtrsim 10^{12} M_{\odot}$), AGN feedback is thought to be the dominant mode \citep[e.g.][]{Croton+2006}, while in less massive galaxies ($M_{\rm halo} \lesssim 10^{12} M_{\odot}$) stellar feedback is most important \citep[e.g.][]{Hopkins+2012, Li+2018, Wang+2018}. 

Present day massive elliptical galaxies (like the ones studied in this paper) grow both via in-situ star formation as well as from ex-situ growth via mergers with lower mass galaxies. The in-situ and ex-situ components will dominate the light profiles of massive galaxies in different radial ranges (Figure \ref{fig:toy_profiles}); in-situ mass is expected to dominate the center of the galaxy and the ex-situ mass becomes dominant at larger radii, though this transition radius moves inward at higher masses \citep{Rodriguez-Gomez+2016, Pillepich+2018b:TNG}. 

AGN feedback limits the growth from in-situ star formation and thus impacts the inner regions of massive galaxies \citep{Martizzi+2014, Peirani+2017}. Stellar feedback is dominant in regulating the mass of the satellites that are accreted to the central galaxy and that deposit stellar mass in the outskirts, thus affecting the outer ex-situ component of the central galaxy's stellar mass profile. By $z=0$, in massive galaxies like the ones in our sample ($M_{\star} > 10^{11.2} M_{\odot}$), more than 40\%  (and up to 80\% in the most massive systems) of the stellar mass comes from mergers (i.e. from their ex-situ component). This has been shown to be the case in Illustris and IllustrisTNG. \citep[e.g.][]{Rodriguez-Gomez+2015, Rodriguez-Gomez+2016, Pillepich+2018b:TNG, Tacchella+2019}. Furthermore, in Illustris, independent of galaxy mass, about half of that ex-situ mass will come from major mergers (1:4 mass ratio), $\sim 40\%$ from minor and very minor mergers, and the remaining $\sim10\%$ is stripped from surviving galaxies \citep{Rodriguez-Gomez+2016}.

In the original Illustris, comparisons with cosmic star formation density measurements compiled from various multi-wavelength surveys \citep{Behroozi&Wechsler&Conroy2013} suggest that the AGN feedback recipe does not quench star-formation sufficiently at low redshift \citep{Vogelsberger+2014:Illustris}. Particularly, in massive halos, the radio-mode feedback, which outputs kinetic energy to mimic the effects of relativistic radio jets in X-ray cavities (``bubble" mode), is not effective at halting star formation. For this reason, in IllustrisTNG the implementation of low and high accretion feedback modes was modified, and the bubble mode was replaced with a kinetic mode that acts at the very center of galaxies.  Indeed, Figure \ref{fig:SHMR+SMFs} shows that TNG100 SMF is improved compared to Illustris \citep[also see][]{Pillepich+2018a, Pillepich+2018b:TNG}. At the lower mass end ($M_{\star} \lesssim 10^{10} M_{\odot}$), this improvement is mostly due to the changes in the implementation of stellar-driven winds, while at higher masses it is due to the modifications to the AGN feedback, both of which suppress the SMF at their respective mass scales. Our new measurements create a new opportunity to yield qualitative new insight into the physical nature of feedback processes. In the following, we discuss how our findings can be interpreted in terms of feedback models. We focus on the results from TNG100 and propose two possible solutions to the relatively mild (dis)agreements between TNG100 and HSC galaxies:

\begin{enumerate}
    \item \textbf{In centrals}: AGN feedback in central galaxies needs to be adjusted to allow for a higher in-situ fraction in lower mass halos ($M_{\rm vir}\sim10^{13}$ M$_{\odot}$) while preventing an excess in higher mass halos ($M_{\rm vir}\sim10^{14}$ M$_{\odot}$).
    
    \item \textbf{In satellites that will eventually merge with centrals in dark matter halos with $M_{\rm vir}\sim10^{14}$ M$_{\odot}$}: feedback from stars and/or AGN in lower mass galaxies ($M_{*}\sim10^{9-10}$ M$_{\odot}$) needs to be modified to suppress their mass growth at early times. This would reduce both the inner regions and the stellar halos for central galaxies of dark matter halos with $M_{\rm vir}=10^{14}$ M$_{\odot}$, bringing TNG100 into closer agreement with HSC.
\end{enumerate}

\subsubsection{In-situ component}

Previous work has demonstrated the significant impact AGN feedback can have on the density profiles of massive elliptical galaxies through its in-situ component \citep{Fan+2008, Peirani+2008, Duffy+2010, Martizzi+2012a, Martizzi+2012b, Peirani+2017, Wang+2019}. Once the galaxy is massive enough and the AGN turns on, radiation and winds from the accretion process at the galactic nucleus will be emitted outward, injecting energy into the galaxy and preventing the cooling of gas to allow for star formation \citep[e.g.][]{Ishibashi&Fabian2012}. When the AGN feedback is absent or not efficient enough at quenching star formation at these masses, there is an overcooling in the central region of these galaxies resulting in increased star-formation and excess mass in the inner regions of the galaxy \citep{Martizzi+2014, Peirani+2017}. Figure \ref{fig:toy_profiles} shows a schematic of how this might affect the stellar mass profiles of massive galaxies. AGN feedback that is not efficient at quenching star formation will result in overcooling and a prominent in-situ component. More efficient AGN feedback would limit star formation and would result in a lower in-situ component.

The inner density profiles of galaxies from HSC can be used as observational constraints on the strength of AGN feedback in central galaxies. In doing so, however, it is important to  keep in mind that towards the highest galaxy masses ($M_{*}\gtrsim10^{11.5}$ M$_{\odot}$), the ex-situ stellar mass becomes the dominant component at progressively smaller galactocentric distances (see next section). The effect of weak to no-AGN feedback is revealed in our comparison with the Ponos simulations (\S \ref{sec:ponos_results}). As in \citet{Martizzi+2014}, the lack of AGN feedback resulted in galaxies that are too massive, particularly in their interior region ($r\lesssim20$ kpc), compared to observations. On larger scales, though, there is good agreement with HSC/Illustris/TNG100. This is interesting as it suggests that the inner profiles of massive galaxies contain information about the physical scale of the impact of AGN feedback on in-situ star formation. The energy injected by the AGN can provide a mechanism for regulating star formation, either by preventing the cooling of gas or by expelling gas from the central regions of galaxies. The details of how and when AGN feedback occurs and its impact on galaxy evolution remain uncertain from both an observational and theoretical point of view. Our analysis allows us to estimate the galaxy scale over which the deposition of energy and momentum via outflows and radiation from the AGN affects the formation of stars in the host galaxy. The comparison shown in Figure \ref{fig:ponos} would suggest that AGN feedback is limiting star formation in massive galaxies at roughly $r\lesssim$ 20 kpc.

Turning now to TNG100, Figure \ref{fig:SHMR+SMFs} displays the $M_{*}^{10}-M_{\rm vir}$ relation and reveals that TNG100 has a steeper relation than HSC with $M_{*}^{10}$ being lower than HSC at $M_{\rm vir}=10^{13}$ M$_{\odot}$ and $M_{*}^{10}$ being higher than HSC at $M_{\rm vir}=10^{14}$ M$_{\odot}$. This could be an indication that TNG100 is too efficient at quenching star formation at $M_{\rm vir}=10^{13}$ M$_{\odot}$ but not efficient enough at $M_{\rm vir}=10^{14}$ M$_{\odot}$. We will study this possibility further in future work by using information about the origin of the stellar particles in question (i.e. whether they were formed in-situ or ex-situ; Chowdhury et al. in preparation).

\subsubsection{Ex-situ component}

Massive galaxies experience significant growth via merging. This growth can significantly impact the stellar mass profiles of massive galaxies given that stars that were formed in-situ are generally closest to the center of the galaxy, followed by stars accreted in major mergers, minor mergers, very minor mergers, and finally stars that were stripped from surviving galaxies \citep[e.g.][]{Rodriguez-Gomez+2015, Rodriguez-Gomez+2016, Pillepich+2018b:TNG, Tacchella+2019}. Therefore, we can appeal to processes that add ex-situ stars to a galaxy to help explain what we see in the outer regions of a galaxy. Overly massive satellites merging with the central may explain the cases in which Illustris and TNG100 galaxies have more extended outskirts compared to HSC. We typically observe this in the most massive halo mass bin.

The most massive galaxies in our sample live in dark matter halos of masses $ \log(M_{\rm vir}/M_{\odot})\gtrsim 14$. In TNG100 (and TNG300), 90\% of the total ex-situ mass of central galaxies (including the ICL component) in halos of these masses is built from progenitors of masses $\log(M_{\star}/M_{\odot})\gtrsim 9-10$ \citep{Pillepich+2018b:TNG}. Most of that ex-situ growth happens through mergers at $z\lesssim1$ \citep{Genel+2018}. While TNG100 has much better agreement with observations than Illustris, there is still a noted excess of lower mass ($\log M_\star/M_{\odot} \lesssim 10$) galaxies at $z\sim 1$ compared to observations \citep[e.g. Figure 14 in][]{Pillepich+2018b:TNG}. It is therefore possible that the excess outer mass we see in the highest halo mass bin in Illustris/TNG100 galaxies compared to HSC is related to an excess ex-situ component built up from overly massive satellite galaxies.

\subsection{Artificial Stripping and Disruption}

Another possibility is that there are non-physical effects in simulations contributing to the excess buildup of mass in the outskirts of galaxies. For example, \citealt{van-den-Bosch2017,van-den-Bosch+2017,van-den-Bosch+2018} have shown that at the numerical resolutions typical of large-volume cosmological simulations, DM subhaloes may undergo over-disruption and over-merging with their central hosts because of numerical effects. Admitting that those studies focus on idealized orbits and in the absence of gas and stars, they nevertheless inspire a critical approach for the interpretation of our findings. As the complete disruption of satellites is not of relevance for this analysis, we instead discuss two hypothetical scenarios: the first is when stellar mass is removed from satellites which should in reality remain gravitationally-bound to satellites; the second is when stellar mass is stripped in the right amount from satellites but deposited in the wrong location (too early or too late) along their orbits.

In the first scenario, satellite galaxies are being artificially over-stripped and this stripped mass adds to the stellar halo (whereas in reality this mass should still be counted to be part of the satellite). 
\citealt{van-den-Bosch+2018} show, albeit in idealized configurations, that, as subhaloes are stripped of their DM, the first 99\% of their mass loss is accurately captured also at the numerical resolutions relevant for this paper. This may indicate that the largest majority of the stripped material (in this case stellar mass) is stripped in a quantitatively consistent manner in simulations like Illustris, IllustrisTNG, or Ponos as in extremely high-resolution controlled experiments, i.e that the lion's share of the stellar mass in the outskirts of massive galaxies should be numerically converged when stripping is concerned (see also Lovell et al. in preparation).

In the second scenario, mass is being stripped in the right amounts, but deposited in the wrong location (e.g. earlier along orbits). Presumably, this would preferentially deposit mass in the outskirts of centrals (at tens or even hundreds of kpc from the center), leading to an excess of mass at large radii and a deficit of mass at small radii, \textit{but the total mass would remain unchanged}. In this paper, we find that in our highest mass bin ($M_{*}\sim10^{14}$ M$_{\odot}$), galaxies in TNG100 are overly massive in total stellar mass (i.e. within 100kpc) compared to HSC (e.g. Figure \ref{fig:SHMR+SMFs}). Hence, we believe that this is unlikely to explain the differences between TNG100 and HSC.

\subsection{Uncertainties in Mass Measurements}
It is also important to take into account the assumptions made in measuring masses from our observed profiles. To convert from surface brightness to mass density, we assume a constant $M/L$ ratio for the entire galaxy. \citet{Huang+2018b} discuss why this assumption is justified given that our sample of massive ellipticals is dominated by old stellar populations and known to have only shallow color gradients (Appendix C in \citealt{Huang+2018b}). Nonetheless, there is a known shallow but negative $M/L$ gradient driven by the metallicity gradient in these galaxies \citep[e.g.][ Chowdhury et al. in preparation]{LaBarbera+2012, D'Souza+2015}. If we instead assume a negative $M/L$ gradient, that would result in decreased masses in the outer regions of our observed HSC galaxies. This would typically result in an even larger discrepancy with the simulated galaxies.

We also consider the limitations of our 1D profile measurement approach. If instead we considered a 2D functional form to approximate the distributions of light (and mass), we would be able to use multiple components to take into account variations in the isophotal shape. We could also integrate it to infinity to derive a true ``total'' mass. We could also take seeing into account. However, we would be limited by the fact that we do not know the perfect functional form (or combination of components) to describe the outskirts of massive galaxies. Multiple-Sersic components can be promising, but internal degeneracy and the lack of clear physical meaning are the downside. These will be explored in a future work (Ardila et al. in preparation).

\section{Summary and Conclusions}\label{sec:conclusions}
In this paper, we perform a consistent comparison of the masses and mass profiles of massive ($M_\star > 10^{11.4}M_\odot$) central galaxies at $z\sim0.4$ from deep HSC observations and from the Illustris, TNG100, and Ponos simulations. HSC weak lensing measurements further enable comparisons at fixed halo mass using a tight relation with with $M_{\star}^{\mathrm{10}}$, and $M_{\star}^{\mathrm{100}}$ from the \texttt{ASAP} empirical model. We measure the stellar mass density profiles, the galaxy stellar mass functions, and the stellar-to-halo mass relations using different operational definitions of stellar masses for massive galaxies in these simulations and compare with measurements from the sample of \citet{Huang+2018b}. Stellar profiles are measured by drawing concentric elliptical isophotes with a fixed ellipticity on the target galaxy, at given radii along the semi-major axis. In the simulations we use stellar mass maps and perform measurements in a way that closely mimics the HSC methodology. Our main results are summarized below.

\begin{itemize}
    \item The direct comparison of galaxy light profiles avoids having to define ``galaxy mass" as a single number and can account for surface brightness limits in observations. The full mass profile also contains more information than mass at any fixed radius. This is shown in Figure \ref{fig:SHMR+SMFs} which compares the galaxy SMF and the SHMR for various aperture definitions of stellar mass. It is clear that different simulations match HSC data better at different radii and in different mass bins. Understanding the radial ranges over which simulations and observations agree/disagree provides important clues about the assembly history and feedback processes that have shaped massive galaxies. 
    
    \item We also compare the weak lensing signals of galaxies from HSC with Illustris/TNG100 (Figure \ref{fig:weak-lensing}). This comparison, together with the \texttt{ASAP} model developed in \citet{Huang+2019}, enables us to perform comparisons at fixed halo mass. The ability to control for halo mass is a key component of this paper as it enables us to draw more informative conclusions about the strength and impact of feedback at different halo mass scales.
    
    \item Both Illustris and TNG100 display overall good agreement with the mass profiles between a few and 100 kpc of massive central galaxies in HSC, albeit with some interesting differences (Figure \ref{fig:matched_by_halo_mass}). Generally, central galaxies in both Illustris and TN100 exhibit steeper stellar-to-halo mass relations, both within 10 and 100 kpc apertures (Figure \ref{fig:SHMR+SMFs}). This disagreement against the HSC observational inferences are more severe in the innermost regions ($r<10$kpc), at higher masses ($M_\star \sim 10^{14}M_\odot$), and in Illustris than in TNG100. In TNG100, the stellar mass distributions of galaxies in $M_\star \sim 10^{13.5}M_\odot$ halos are in excellent agreement with observations, while those in $M_\star \sim 10^{13}M_\odot$ halos have a deficit of stellar mass (by 0.05-0.15 dex, depending on radius) and those in halos of $M_\star \sim 10^{14}M_\odot$ have an excess (by 0.15-0.30 dex). In fact, TNG100 returns an outer stellar mass ($M_{*}^{100}-M_{*}^{10}$) that is consistent with the SMDPL+ASAP results to better than 0.12 dex across the halo mass range studied here.
    
    \item We interpret our results by assuming the two-phase formation scenario of massive galaxies in which the inner regions of $M_\star \sim 10^{11}M_\odot$ galaxies are dominated by in-situ mass and the outer regions are dominated by ex-situ mass (Figure \ref{fig:toy_profiles}). Given that both the inner and outer stellar masses show offsets in some halo mass bins in both Illustris and TNG100, the culprit for the (dis)agreement may be found in both in-situ and ex-situ stellar mass processes. The steeper $M_{*}^{10}-M_{\rm vir}$ relation in TNG100 than in HSC (TNG100 $M_{*}^{10}$ is lower than HSC at $M_{\rm vir}\sim10^{13}M_\odot$ and higher at $M_{\rm vir}\sim10^{14}M_\odot$; Figure \ref{fig:SHMR+SMFs}) may be signaling that quenching in the TNG100 model is too efficient at $M_{\rm vir}\sim10^{13}M_\odot$ but not efficient enough for the stars that end up assembling in $M_{\rm vir}\sim10^{14}$ halos. On the other hand, central galaxies in $M_{\rm vir}\sim10^{14}M_\odot$ halos display too much outer mass in TNG100 compared to HSC (albeit by only $\sim0.12$ dex in $M_{*}^{100}-M_{*}^{10}$). We hypothesize that feedback from stars and/or AGN needs to be adjusted to suppress the mass growth at early times of satellites that build up the stellar halo ($M_\star \sim 10^{9-10} M_{\odot}$). Resolution effects could also be at play and warrant further investigation.

    \item Galaxies in the zoom-in Ponos simulation, which does not implement AGN feedback, display a substantial excess mass of $\sim 0.5$ dex at $r<30$ kpc compared to HSC galaxies of similar halo mass (though a comparable slope and amplitude for the rest of the profile). This is indicative of over-cooling and excess star formation in the central regions due to the lack of regulation from the central black hole in the Ponos simulations Figure \ref{fig:ponos}). The comparison between Ponos and HSC suggests that the physical scale over which the central AGN limits star formation is $r\lesssim20$ kpc.
    
    \item We have also performed a number of tests to validate our results. In particular, we showed the choice of method used to mask neighbouring satellite galaxies does not impact our measurements. This is due to the iterative 3$\sigma$--clipping that we apply along each isophote \textit{prior} to measuring the mean intensity. Even when a satellite mask is not applied, the impact of satellites represents less than a 2\% effect for 95\% of our sample at 150 kpc.
    
    \item We provide median surface stellar mass density profiles and weak-lensing $\Delta\Sigma$ profiles for the HSC galaxies (\url{https://github.com/f-ardila/HSC_vs_hydro-paper}) to facilitate comparisons with other simulations.
\end{itemize}

In the era of large surveys that are wide, deep, and that have lensing capability (HSC, LSST, Euclid, WFIRST), joint comparisons between weak lensing and galaxy stellar profiles offer a powerful method for comparing simulations with observations. In particular, this combination provides a direct test of whether simulations build and deposit galaxy mass in the correct dark matter halos and therefore constrains the physics of feedback and galaxy growth.

\section*{Acknowledgements}

FA is supported by the National Science Foundation Graduate Research Fellowship Program under Grant No. DGE-1842400.
This research was supported in part by the National Science Foundation under Grant No. NSF PHY-1748958. 
This material is based upon work supported by the National Science Foundation under Grant No. 1714610. This research was also supported in part by National Science Foundation under Grant No. NSF PHY11-25915. AL acknowledges support from the David and Lucille Packard foundation, and from the Alfred P. Sloan Foundation. We acknowledge use of the lux supercomputer at UC Santa Cruz, funded by NSF MRI grant AST 1828315. PM acknowledges a NASA contract supporting the WFIRST-EXPO Science Investigation Team (15-WFIRST15-0004), administered by the 
GSFC. Work done at Argonne National Laboratory was supported under the DOE contract DE-AC02-06CH11357.

Based [in part] on data collected at the Subaru Telescope and retrieved from the HSC data archive system, which is operated by the Subaru Telescope and Astronomy Data Center at National Astronomical Observatory of Japan.

The Hyper Suprime-Cam (HSC) collaboration includes the astronomical communities of Japan and Taiwan, and Princeton University. The HSC instrumentation and software were developed by the National Astronomical Observatory of Japan (NAOJ), the Kavli Institute for the Physics and Mathematics of the Universe (Kavli IPMU), the University of Tokyo, the High Energy Accelerator Research Organization (KEK), the Academia Sinica Institute for Astronomy and Astrophysics in Taiwan (ASIAA), and Princeton University. Funding was contributed by the FIRST program from Japanese Cabinet Office, the Ministry of Education, Culture, Sports, Science and Technology (MEXT), the Japan Society for the Promotion of Science (JSPS), Japan Science and Technology Agency (JST), the Toray Science Foundation, NAOJ, Kavli IPMU, KEK, ASIAA, and Princeton University.

The Pan-STARRS1 Surveys (PS1) have been made possible through contributions of the Institute for Astronomy, the University of Hawaii, the Pan-STARRS Project Office, the Max-Planck Society and its participating institutes, the Max Planck Institute for Astronomy, Heidelberg and the Max Planck Institute for Extraterrestrial Physics, Garching, The Johns Hopkins University, Durham University, the University of Edinburgh, Queen's University Belfast, the Harvard-Smithsonian Center for Astrophysics, the Las Cumbres Observatory Global Telescope Network Incorporated, the National Central University of Taiwan, the Space Telescope Science Institute, the National Aeronautics and Space Administration under Grant No. NNX08AR22G issued through the Planetary Science Division of the NASA Science Mission Directorate, the National Science Foundation under Grant No. AST-1238877, the University of Maryland, and Eotvos Lorand University (ELTE).

This paper makes use of software developed for the Large Synoptic Survey Telescope. We thank the LSST Project for making their code available as free software at \url{http://dm.lsst.org}.

The following software and programming languages made this research possible: \texttt{PYTHON} (v2.7); \texttt{ASTROPY} \citep[v2.0;][]{astropy2013, astropy2018}, a community-developed core \texttt{PYTHON} package for Astronomy; Source Extraction and Photometry in Python \citep[\texttt{SEP} v1.0;][]{sep2015}; \texttt{Halotools} \citep[v0.7;][]{halotools}; COsmology, haLO and large-Scale StrUcture toolS \citep[\texttt{COLOSSUS;}][]{colossus2018}. 

FA would like to express deep gratitude to his grandfather, Andr\'es Rosselli Quijano, for all the knowledge and passion he passed on, as well as the inspiration to pursue this subject area.

\section*{Data Availability}
The observational HSC data underlying this article are publicly available at \url{https://github.com/f-ardila/HSC_vs_hydro-paper}. Illustris/TNG simulation data are publicly available at \url{https://www.tng-project.org/data/}.
\bibliographystyle{mnras}
\bibliography{this_paper}

\begin{thebibliography}{}
\makeatletter
\relax
\def\mn@urlcharsother{\let\do\@makeother \do\$\do\&\do\#\do\^\do\_\do\%\do\~}
\def\mn@doi{\begingroup\mn@urlcharsother \@ifnextchar [ {\mn@doi@}
  {\mn@doi@[]}}
\def\mn@doi@[#1]#2{\def\@tempa{#1}\ifx\@tempa\@empty \href
  {http://dx.doi.org/#2} {doi:#2}\else \href {http://dx.doi.org/#2} {#1}\fi
  \endgroup}
\def\mn@eprint#1#2{\mn@eprint@#1:#2::\@nil}
\def\mn@eprint@arXiv#1{\href {http://arxiv.org/abs/#1} {{\tt arXiv:#1}}}
\def\mn@eprint@dblp#1{\href {http://dblp.uni-trier.de/rec/bibtex/#1.xml}
  {dblp:#1}}
\def\mn@eprint@#1:#2:#3:#4\@nil{\def\@tempa {#1}\def\@tempb {#2}\def\@tempc
  {#3}\ifx \@tempc \@empty \let \@tempc \@tempb \let \@tempb \@tempa \fi \ifx
  \@tempb \@empty \def\@tempb {arXiv}\fi \@ifundefined
  {mn@eprint@\@tempb}{\@tempb:\@tempc}{\expandafter \expandafter \csname
  mn@eprint@\@tempb\endcsname \expandafter{\@tempc}}}

\bibitem[\protect\citeauthoryear{{Aihara} et~al.,}{{Aihara}
  et~al.}{2018a}]{Aihara+2018}
{Aihara} H.,  et~al., 2018a, \mn@doi [\pasj] {10.1093/pasj/psx066}, \href
  {https://ui.adsabs.harvard.edu/abs/2018PASJ...70S...4A} {70, S4}

\bibitem[\protect\citeauthoryear{{Aihara} et~al.,}{{Aihara}
  et~al.}{2018b}]{Aihara+2017:SSP}
{Aihara} H.,  et~al., 2018b, \mn@doi [\pasj] {10.1093/pasj/psx081}, \href
  {https://ui.adsabs.harvard.edu/abs/2018PASJ...70S...8A} {70, S8}

\bibitem[\protect\citeauthoryear{{Astropy Collaboration} et~al.,}{{Astropy
  Collaboration} et~al.}{2013}]{astropy2013}
{Astropy Collaboration} et~al., 2013, \mn@doi [\aap]
  {10.1051/0004-6361/201322068}, \href
  {http://adsabs.harvard.edu/abs/2013A%26A...558A..33A} {558, A33}

\bibitem[\protect\citeauthoryear{{Astropy Collaboration} et~al.,}{{Astropy
  Collaboration} et~al.}{2018}]{astropy2018}
{Astropy Collaboration} et~al., 2018, \mn@doi [\aj] {10.3847/1538-3881/aabc4f},
  \href {http://adsabs.harvard.edu/abs/2018AJ....156..123A} {156, 123}

\bibitem[\protect\citeauthoryear{{Baldry}, {Glazebrook}  \& {Driver}}{{Baldry}
  et~al.}{2008}]{Baldry+2008}
{Baldry} I.~K.,  {Glazebrook} K.,   {Driver} S.~P.,  2008, \mn@doi [\mnras]
  {10.1111/j.1365-2966.2008.13348.x}, \href
  {https://ui.adsabs.harvard.edu/abs/2008MNRAS.388..945B} {388, 945}

\bibitem[\protect\citeauthoryear{Barbary, Boone  \& Deil}{Barbary
  et~al.}{2015}]{sep2015}
Barbary K.,  Boone K.,   Deil C.,  2015, sep: v0.3.0,
  \mn@doi{10.5281/zenodo.15669}, \url {http://dx.doi.org/10.5281/zenodo.15669}

\bibitem[\protect\citeauthoryear{{Beckmann} et~al.,}{{Beckmann}
  et~al.}{2017}]{Beckmann+2017}
{Beckmann} R.~S.,  et~al., 2017, \mn@doi [\mnras] {10.1093/mnras/stx1831},
  \href {https://ui.adsabs.harvard.edu/abs/2017MNRAS.472..949B} {472, 949}

\bibitem[\protect\citeauthoryear{{Behroozi}, {Wechsler}  \&
  {Conroy}}{{Behroozi} et~al.}{2013}]{Behroozi&Wechsler&Conroy2013}
{Behroozi} P.~S.,  {Wechsler} R.~H.,   {Conroy} C.,  2013, \mn@doi [\apj]
  {10.1088/0004-637X/770/1/57}, \href
  {https://ui.adsabs.harvard.edu/abs/2013ApJ...770...57B} {770, 57}

\bibitem[\protect\citeauthoryear{{Behroozi}, {Wechsler}, {Hearin}  \&
  {Conroy}}{{Behroozi} et~al.}{2019}]{Behroozi+2019}
{Behroozi} P.,  {Wechsler} R.~H.,  {Hearin} A.~P.,   {Conroy} C.,  2019,
  \mn@doi [\mnras] {10.1093/mnras/stz1182}, \href
  {https://ui.adsabs.harvard.edu/abs/2019MNRAS.488.3143B} {488, 3143}

\bibitem[\protect\citeauthoryear{{Benson}, {Bower}, {Frenk}, {Lacey}, {Baugh}
  \& {Cole}}{{Benson} et~al.}{2003}]{Benson+2003}
{Benson} A.~J.,  {Bower} R.~G.,  {Frenk} C.~S.,  {Lacey} C.~G.,  {Baugh} C.~M.,
    {Cole} S.,  2003, \mn@doi [\apj] {10.1086/379160}, \href
  {https://ui.adsabs.harvard.edu/abs/2003ApJ...599...38B} {599, 38}

\bibitem[\protect\citeauthoryear{{Bernardi}, {Meert}, {Sheth}, {Vikram},
  {Huertas-Company}, {Mei}  \& {Shankar}}{{Bernardi}
  et~al.}{2013}]{Bernardi+2013}
{Bernardi} M.,  {Meert} A.,  {Sheth} R.~K.,  {Vikram} V.,  {Huertas-Company}
  M.,  {Mei} S.,   {Shankar} F.,  2013, \mn@doi [\mnras]
  {10.1093/mnras/stt1607}, \href
  {https://ui.adsabs.harvard.edu/abs/2013MNRAS.436..697B} {436, 697}

\bibitem[\protect\citeauthoryear{{Bertin} \& {Arnouts}}{{Bertin} \&
  {Arnouts}}{1996}]{sextractor}
{Bertin} E.,  {Arnouts} S.,  1996, \mn@doi [\aaps] {10.1051/aas:1996164}, \href
  {https://ui.adsabs.harvard.edu/abs/1996A%26AS..117..393B} {117, 393}

\bibitem[\protect\citeauthoryear{{Borgani} \& {Kravtsov}}{{Borgani} \&
  {Kravtsov}}{2011}]{Borgani&Kravtsov2011:review_cluster_sims}
{Borgani} S.,  {Kravtsov} A.,  2011, \mn@doi [Advanced Science Letters]
  {10.1166/asl.2011.1209}, \href
  {https://ui.adsabs.harvard.edu/abs/2011ASL.....4..204B} {4, 204}

\bibitem[\protect\citeauthoryear{{Bryan} \& {Norman}}{{Bryan} \&
  {Norman}}{1998}]{Bryan&Norman1998}
{Bryan} G.~L.,  {Norman} M.~L.,  1998, \mn@doi [\apj] {10.1086/305262}, \href
  {https://ui.adsabs.harvard.edu/abs/1998ApJ...495...80B} {495, 80}

\bibitem[\protect\citeauthoryear{{Carollo}, {Danziger}  \& {Buson}}{{Carollo}
  et~al.}{1993}]{Carollo+1993}
{Carollo} C.~M.,  {Danziger} I.~J.,   {Buson} L.,  1993, \mn@doi [\mnras]
  {10.1093/mnras/265.3.553}, \href
  {https://ui.adsabs.harvard.edu/abs/1993MNRAS.265..553C} {265, 553}

\bibitem[\protect\citeauthoryear{{Croton} et~al.,}{{Croton}
  et~al.}{2006}]{Croton+2006}
{Croton} D.~J.,  et~al., 2006, \mn@doi [\mnras]
  {10.1111/j.1365-2966.2005.09675.x}, \href
  {https://ui.adsabs.harvard.edu/abs/2006MNRAS.365...11C} {365, 11}

\bibitem[\protect\citeauthoryear{{D'Souza}, {Kauffman}, {Wang}  \&
  {Vegetti}}{{D'Souza} et~al.}{2014}]{D'Souza+2014}
{D'Souza} R.,  {Kauffman} G.,  {Wang} J.,   {Vegetti} S.,  2014, \mn@doi
  [\mnras] {10.1093/mnras/stu1194}, \href
  {https://ui.adsabs.harvard.edu/abs/2014MNRAS.443.1433D} {443, 1433}

\bibitem[\protect\citeauthoryear{{D'Souza}, {Vegetti}  \&
  {Kauffmann}}{{D'Souza} et~al.}{2015}]{D'Souza+2015}
{D'Souza} R.,  {Vegetti} S.,   {Kauffmann} G.,  2015, \mn@doi [\mnras]
  {10.1093/mnras/stv2234}, \href
  {https://ui.adsabs.harvard.edu/abs/2015MNRAS.454.4027D} {454, 4027}

\bibitem[\protect\citeauthoryear{{Davies}, {Sadler}  \& {Peletier}}{{Davies}
  et~al.}{1993}]{Davies+1993}
{Davies} R.~L.,  {Sadler} E.~M.,   {Peletier} R.~F.,  1993, \mn@doi [\mnras]
  {10.1093/mnras/262.3.650}, \href
  {https://ui.adsabs.harvard.edu/abs/1993MNRAS.262..650D} {262, 650}

\bibitem[\protect\citeauthoryear{{Diemer}}{{Diemer}}{2018}]{colossus2018}
{Diemer} B.,  2018, \mn@doi [\apjs] {10.3847/1538-4365/aaee8c}, \href
  {https://ui.adsabs.harvard.edu/abs/2018ApJS..239...35D} {239, 35}

\bibitem[\protect\citeauthoryear{{Donnari} et~al.,}{{Donnari}
  et~al.}{2019}]{Donnari+2019}
{Donnari} M.,  et~al., 2019, \mn@doi [\mnras] {10.1093/mnras/stz712}, \href
  {https://ui.adsabs.harvard.edu/abs/2019MNRAS.485.4817D} {485, 4817}

\bibitem[\protect\citeauthoryear{{Dubois}, {Peirani}, {Pichon}, {Devriendt},
  {Gavazzi}, {Welker}  \& {Volonteri}}{{Dubois} et~al.}{2016}]{Dubois+2016}
{Dubois} Y.,  {Peirani} S.,  {Pichon} C.,  {Devriendt} J.,  {Gavazzi} R.,
  {Welker} C.,   {Volonteri} M.,  2016, \mn@doi [\mnras]
  {10.1093/mnras/stw2265}, \href
  {https://ui.adsabs.harvard.edu/abs/2016MNRAS.463.3948D} {463, 3948}

\bibitem[\protect\citeauthoryear{{Duffy}, {Schaye}, {Kay}, {Dalla Vecchia},
  {Battye}  \& {Booth}}{{Duffy} et~al.}{2010}]{Duffy+2010}
{Duffy} A.~R.,  {Schaye} J.,  {Kay} S.~T.,  {Dalla Vecchia} C.,  {Battye}
  R.~A.,   {Booth} C.~M.,  2010, \mn@doi [\mnras]
  {10.1111/j.1365-2966.2010.16613.x}, \href
  {https://ui.adsabs.harvard.edu/abs/2010MNRAS.405.2161D} {405, 2161}

\bibitem[\protect\citeauthoryear{{Fan}, {Lapi}, {De Zotti}  \& {Danese}}{{Fan}
  et~al.}{2008}]{Fan+2008}
{Fan} L.,  {Lapi} A.,  {De Zotti} G.,   {Danese} L.,  2008, \mn@doi [\apjl]
  {10.1086/595784}, \href
  {https://ui.adsabs.harvard.edu/abs/2008ApJ...689L.101F} {689, L101}

\bibitem[\protect\citeauthoryear{{Fiacconi}, {Madau}, {Potter}  \&
  {Stadel}}{{Fiacconi} et~al.}{2016}]{Fiacconi+2016}
{Fiacconi} D.,  {Madau} P.,  {Potter} D.,   {Stadel} J.,  2016, \mn@doi [\apj]
  {10.3847/0004-637X/824/2/144}, \href
  {http://adsabs.harvard.edu/abs/2016ApJ...824..144F} {824, 144}

\bibitem[\protect\citeauthoryear{{Fiacconi}, {Mayer}, {Madau}, {Lupi}, {Dotti}
  \& {Haardt}}{{Fiacconi} et~al.}{2017}]{Fiacconi+2017}
{Fiacconi} D.,  {Mayer} L.,  {Madau} P.,  {Lupi} A.,  {Dotti} M.,   {Haardt}
  F.,  2017, \mn@doi [\mnras] {10.1093/mnras/stx335}, \href
  {https://ui.adsabs.harvard.edu/abs/2017MNRAS.467.4080F} {467, 4080}

\bibitem[\protect\citeauthoryear{{Genel} et~al.,}{{Genel}
  et~al.}{2014}]{Genel+2014:Illustris}
{Genel} S.,  et~al., 2014, \mn@doi [\mnras] {10.1093/mnras/stu1654}, \href
  {https://ui.adsabs.harvard.edu/abs/2014MNRAS.445..175G} {445, 175}

\bibitem[\protect\citeauthoryear{{Genel} et~al.,}{{Genel}
  et~al.}{2018}]{Genel+2018}
{Genel} S.,  et~al., 2018, \mn@doi [\mnras] {10.1093/mnras/stx3078}, \href
  {https://ui.adsabs.harvard.edu/abs/2018MNRAS.474.3976G} {474, 3976}

\bibitem[\protect\citeauthoryear{{Hearin} et~al.,}{{Hearin}
  et~al.}{2017}]{halotools}
{Hearin} A.~P.,  et~al., 2017, \mn@doi [\aj] {10.3847/1538-3881/aa859f}, \href
  {https://ui.adsabs.harvard.edu/abs/2017AJ....154..190H} {154, 190}

\bibitem[\protect\citeauthoryear{{Hopkins}, {Quataert}  \& {Murray}}{{Hopkins}
  et~al.}{2012}]{Hopkins+2012}
{Hopkins} P.~F.,  {Quataert} E.,   {Murray} N.,  2012, \mn@doi [\mnras]
  {10.1111/j.1365-2966.2012.20593.x}, \href
  {https://ui.adsabs.harvard.edu/abs/2012MNRAS.421.3522H} {421, 3522}

\bibitem[\protect\citeauthoryear{{Huang} et~al.,}{{Huang}
  et~al.}{2018a}]{Huang+2018c}
{Huang} S.,  et~al., 2018a, preprint, \href
  {http://adsabs.harvard.edu/abs/2018arXiv180302824H} {} (\mn@eprint {arXiv}
  {1803.02824})

\bibitem[\protect\citeauthoryear{{Huang}, {Leauthaud}, {Greene}, {Bundy},
  {Lin}, {Tanaka}, {Miyazaki}  \& {Komiyama}}{{Huang}
  et~al.}{2018b}]{Huang+2018b}
{Huang} S.,  {Leauthaud} A.,  {Greene} J.~E.,  {Bundy} K.,  {Lin} Y.-T.,
  {Tanaka} M.,  {Miyazaki} S.,   {Komiyama} Y.,  2018b, \mn@doi [\mnras]
  {10.1093/mnras/stx3200}, \href
  {http://adsabs.harvard.edu/abs/2018MNRAS.475.3348H} {475, 3348}

\bibitem[\protect\citeauthoryear{{Huang} et~al.,}{{Huang}
  et~al.}{2019}]{Huang+2019}
{Huang} S.,  et~al., 2019, \mn@doi [\mnras] {10.1093/mnras/stz3314}, \href
  {https://ui.adsabs.harvard.edu/abs/2019MNRAS.tmp.2990H} {p.~2990}

\bibitem[\protect\citeauthoryear{{Ishibashi} \& {Fabian}}{{Ishibashi} \&
  {Fabian}}{2012}]{Ishibashi&Fabian2012}
{Ishibashi} W.,  {Fabian} A.~C.,  2012, \mn@doi [\mnras]
  {10.1111/j.1365-2966.2012.22074.x}, \href
  {https://ui.adsabs.harvard.edu/abs/2012MNRAS.427.2998I} {427, 2998}

\bibitem[\protect\citeauthoryear{{Keller}, {Wadsley}, {Benincasa}  \&
  {Couchman}}{{Keller} et~al.}{2014}]{Keller+2014}
{Keller} B.~W.,  {Wadsley} J.,  {Benincasa} S.~M.,   {Couchman} H.~M.~P.,
  2014, \mn@doi [\mnras] {10.1093/mnras/stu1058}, \href
  {https://ui.adsabs.harvard.edu/abs/2014MNRAS.442.3013K} {442, 3013}

\bibitem[\protect\citeauthoryear{{Keller}, {Wadsley}  \& {Couchman}}{{Keller}
  et~al.}{2015}]{Keller+2015}
{Keller} B.~W.,  {Wadsley} J.,   {Couchman} H.~M.~P.,  2015, \mn@doi [\mnras]
  {10.1093/mnras/stv1789}, \href
  {https://ui.adsabs.harvard.edu/abs/2015MNRAS.453.3499K} {453, 3499}

\bibitem[\protect\citeauthoryear{{Klypin}, {Yepes}, {Gottl{\"o}ber}, {Prada}
  \& {He{\ss}}}{{Klypin} et~al.}{2016}]{Klypin+2016:MDPL2}
{Klypin} A.,  {Yepes} G.,  {Gottl{\"o}ber} S.,  {Prada} F.,   {He{\ss}} S.,
  2016, \mn@doi [\mnras] {10.1093/mnras/stw248}, \href
  {https://ui.adsabs.harvard.edu/abs/2016MNRAS.457.4340K} {457, 4340}

\bibitem[\protect\citeauthoryear{{La Barbera}, {De Carvalho}, {De La Rosa},
  {Gal}, {Swindle}  \& {Lopes}}{{La Barbera} et~al.}{2010}]{LaBarbera+2010}
{La Barbera} F.,  {De Carvalho} R.~R.,  {De La Rosa} I.~G.,  {Gal} R.~R.,
  {Swindle} R.,   {Lopes} P.~A.~A.,  2010, \mn@doi [\aj]
  {10.1088/0004-6256/140/5/1528}, \href
  {https://ui.adsabs.harvard.edu/abs/2010AJ....140.1528L} {140, 1528}

\bibitem[\protect\citeauthoryear{{La Barbera}, {Ferreras}, {de Carvalho},
  {Bruzual}, {Charlot}, {Pasquali}  \& {Merlin}}{{La Barbera}
  et~al.}{2012}]{LaBarbera+2012}
{La Barbera} F.,  {Ferreras} I.,  {de Carvalho} R.~R.,  {Bruzual} G.,
  {Charlot} S.,  {Pasquali} A.,   {Merlin} E.,  2012, \mn@doi [\mnras]
  {10.1111/j.1365-2966.2012.21848.x}, \href
  {https://ui.adsabs.harvard.edu/abs/2012MNRAS.426.2300L} {426, 2300}

\bibitem[\protect\citeauthoryear{{Laigle} et~al.,}{{Laigle}
  et~al.}{2019}]{Laigle+2019}
{Laigle} C.,  et~al., 2019, \mn@doi [\mnras] {10.1093/mnras/stz1054}, \href
  {https://ui.adsabs.harvard.edu/abs/2019MNRAS.486.5104L} {486, 5104}

\bibitem[\protect\citeauthoryear{{Leauthaud} et~al.,}{{Leauthaud}
  et~al.}{2016}]{Leauthaud+2016}
{Leauthaud} A.,  et~al., 2016, \mn@doi [\mnras] {10.1093/mnras/stw117}, \href
  {https://ui.adsabs.harvard.edu/abs/2016MNRAS.457.4021L} {457, 4021}

\bibitem[\protect\citeauthoryear{{Li} et~al.,}{{Li} et~al.}{2018}]{Li+2018}
{Li} Y.-P.,  et~al., 2018, \mn@doi [\apj] {10.3847/1538-4357/aade8b}, \href
  {https://ui.adsabs.harvard.edu/abs/2018ApJ...866...70L} {866, 70}

\bibitem[\protect\citeauthoryear{{Mandelbaum} et~al.,}{{Mandelbaum}
  et~al.}{2018}]{Mandelbaum+2018}
{Mandelbaum} R.,  et~al., 2018, \mn@doi [\pasj] {10.1093/pasj/psx130}, \href
  {http://adsabs.harvard.edu/abs/2018PASJ...70S..25M} {70, S25}

\bibitem[\protect\citeauthoryear{{Marinacci} et~al.,}{{Marinacci}
  et~al.}{2018}]{Marinacci+2018:TNG}
{Marinacci} F.,  et~al., 2018, \mn@doi [\mnras] {10.1093/mnras/sty2206}, \href
  {https://ui.adsabs.harvard.edu/abs/2018MNRAS.480.5113M} {480, 5113}

\bibitem[\protect\citeauthoryear{{Martizzi}, {Teyssier}  \& {Moore}}{{Martizzi}
  et~al.}{2012a}]{Martizzi+2012a}
{Martizzi} D.,  {Teyssier} R.,   {Moore} B.,  2012a, \mn@doi [\mnras]
  {10.1111/j.1365-2966.2011.19950.x}, \href
  {https://ui.adsabs.harvard.edu/abs/2012MNRAS.420.2859M} {420, 2859}

\bibitem[\protect\citeauthoryear{{Martizzi}, {Teyssier}, {Moore}  \&
  {Wentz}}{{Martizzi} et~al.}{2012b}]{Martizzi+2012b}
{Martizzi} D.,  {Teyssier} R.,  {Moore} B.,   {Wentz} T.,  2012b, \mn@doi
  [\mnras] {10.1111/j.1365-2966.2012.20879.x}, \href
  {https://ui.adsabs.harvard.edu/abs/2012MNRAS.422.3081M} {422, 3081}

\bibitem[\protect\citeauthoryear{{Martizzi}, {Jimmy}, {Teyssier}  \&
  {Moore}}{{Martizzi} et~al.}{2014}]{Martizzi+2014}
{Martizzi} D.,  {Jimmy} {Teyssier} R.,   {Moore} B.,  2014, \mn@doi [\mnras]
  {10.1093/mnras/stu1233}, \href
  {https://ui.adsabs.harvard.edu/abs/2014MNRAS.443.1500M} {443, 1500}

\bibitem[\protect\citeauthoryear{{Mayer}, {Tamburello}, {Lupi}, {Keller},
  {Wadsley}  \& {Madau}}{{Mayer} et~al.}{2016}]{Mayer+2016}
{Mayer} L.,  {Tamburello} V.,  {Lupi} A.,  {Keller} B.,  {Wadsley} J.,
  {Madau} P.,  2016, \mn@doi [\apjl] {10.3847/2041-8205/830/1/L13}, \href
  {https://ui.adsabs.harvard.edu/abs/2016ApJ...830L..13M} {830, L13}

\bibitem[\protect\citeauthoryear{Miyazaki et~al.,}{Miyazaki
  et~al.}{2012}]{Miyazaki+2012:HSC}
Miyazaki S.,  et~al., 2012, in McLean I.~S.,  Ramsay S.~K.,   Takami H.,  eds,
  ~ Vol. 8446, Ground-based and Airborne Instrumentation for Astronomy IV.
  SPIE, pp 327 -- 335, \mn@doi{10.1117/12.926844}

\bibitem[\protect\citeauthoryear{{Miyazaki} et~al.,}{{Miyazaki}
  et~al.}{2018}]{Miyazaki+2018}
{Miyazaki} S.,  et~al., 2018, \mn@doi [\pasj] {10.1093/pasj/psx063}, \href
  {https://ui.adsabs.harvard.edu/abs/2018PASJ...70S...1M} {70, S1}

\bibitem[\protect\citeauthoryear{{Muzzin} et~al.,}{{Muzzin}
  et~al.}{2013}]{Muzzin+2013}
{Muzzin} A.,  et~al., 2013, \mn@doi [\apj] {10.1088/0004-637X/777/1/18}, \href
  {https://ui.adsabs.harvard.edu/abs/2013ApJ...777...18M} {777, 18}

\bibitem[\protect\citeauthoryear{{Naiman} et~al.,}{{Naiman}
  et~al.}{2018}]{Naiman+2018:TNG}
{Naiman} J.~P.,  et~al., 2018, \mn@doi [\mnras] {10.1093/mnras/sty618}, \href
  {https://ui.adsabs.harvard.edu/abs/2018MNRAS.477.1206N} {477, 1206}

\bibitem[\protect\citeauthoryear{{Nelson} et~al.,}{{Nelson}
  et~al.}{2015}]{Nelson+2015:illustris}
{Nelson} D.,  et~al., 2015, \mn@doi [Astronomy and Computing]
  {10.1016/j.ascom.2015.09.003}, \href
  {https://ui.adsabs.harvard.edu/abs/2015A&C....13...12N} {13, 12}

\bibitem[\protect\citeauthoryear{{Nelson} et~al.,}{{Nelson}
  et~al.}{2018}]{Nelson+2018:TNG}
{Nelson} D.,  et~al., 2018, \mn@doi [\mnras] {10.1093/mnras/stx3040}, \href
  {https://ui.adsabs.harvard.edu/abs/2018MNRAS.475..624N} {475, 624}

\bibitem[\protect\citeauthoryear{{Nelson} et~al.,}{{Nelson}
  et~al.}{2019}]{Nelson+2019:tng}
{Nelson} D.,  et~al., 2019, \mn@doi [Computational Astrophysics and Cosmology]
  {10.1186/s40668-019-0028-x}, \href
  {https://ui.adsabs.harvard.edu/abs/2019ComAC...6....2N} {6, 2}

\bibitem[\protect\citeauthoryear{{Peirani}, {Kay}  \& {Silk}}{{Peirani}
  et~al.}{2008}]{Peirani+2008}
{Peirani} S.,  {Kay} S.,   {Silk} J.,  2008, \mn@doi [\aap]
  {10.1051/0004-6361:20077956}, \href
  {https://ui.adsabs.harvard.edu/abs/2008A&A...479..123P} {479, 123}

\bibitem[\protect\citeauthoryear{{Peirani} et~al.,}{{Peirani}
  et~al.}{2017}]{Peirani+2017}
{Peirani} S.,  et~al., 2017, \mn@doi [\mnras] {10.1093/mnras/stx2099}, \href
  {https://ui.adsabs.harvard.edu/abs/2017MNRAS.472.2153P} {472, 2153}

\bibitem[\protect\citeauthoryear{{P{\'e}rez-Gonz{\'a}lez}, {Trujillo}, {Barro},
  {Gallego}, {Zamorano}  \& {Conselice}}{{P{\'e}rez-Gonz{\'a}lez}
  et~al.}{2008}]{Perez-Gonzalez+2008}
{P{\'e}rez-Gonz{\'a}lez} P.~G.,  {Trujillo} I.,  {Barro} G.,  {Gallego} J.,
  {Zamorano} J.,   {Conselice} C.~J.,  2008, \mn@doi [\apj] {10.1086/591843},
  \href {http://adsabs.harvard.edu/abs/2008ApJ...687...50P} {687, 50}

\bibitem[\protect\citeauthoryear{{Pillepich} et~al.,}{{Pillepich}
  et~al.}{2018a}]{Pillepich+2018a}
{Pillepich} A.,  et~al., 2018a, \mn@doi [\mnras] {10.1093/mnras/stx2656}, \href
  {https://ui.adsabs.harvard.edu/abs/2018MNRAS.473.4077P} {473, 4077}

\bibitem[\protect\citeauthoryear{{Pillepich} et~al.,}{{Pillepich}
  et~al.}{2018b}]{Pillepich+2018b:TNG}
{Pillepich} A.,  et~al., 2018b, \mn@doi [\mnras] {10.1093/mnras/stx3112}, \href
  {https://ui.adsabs.harvard.edu/abs/2018MNRAS.475..648P} {475, 648}

\bibitem[\protect\citeauthoryear{{Price}, {Kriek}, {Feldmann}, {Quataert},
  {Hopkins}, {Faucher-Gigu{\`e}re}, {Kere{\v s}}  \& {Barro}}{{Price}
  et~al.}{2017}]{Price2017}
{Price} S.~H.,  {Kriek} M.,  {Feldmann} R.,  {Quataert} E.,  {Hopkins} P.~F.,
  {Faucher-Gigu{\`e}re} C.-A.,  {Kere{\v s}} D.,   {Barro} G.,  2017, \mn@doi
  [\apjl] {10.3847/2041-8213/aa7d4b}, \href
  {http://adsabs.harvard.edu/abs/2017ApJ...844L...6P} {844, L6}

\bibitem[\protect\citeauthoryear{{Remus}, {Dolag}  \& {Hoffmann}}{{Remus}
  et~al.}{2017}]{Remus+2017}
{Remus} R.-S.,  {Dolag} K.,   {Hoffmann} T.,  2017, \mn@doi [Galaxies]
  {10.3390/galaxies5030049}, \href
  {https://ui.adsabs.harvard.edu/abs/2017Galax...5...49R} {5, 49}

\bibitem[\protect\citeauthoryear{{Rodriguez-Gomez} et~al.,}{{Rodriguez-Gomez}
  et~al.}{2015}]{Rodriguez-Gomez+2015}
{Rodriguez-Gomez} V.,  et~al., 2015, \mn@doi [\mnras] {10.1093/mnras/stv264},
  \href {http://adsabs.harvard.edu/abs/2015MNRAS.449...49R} {449, 49}

\bibitem[\protect\citeauthoryear{{Rodriguez-Gomez} et~al.,}{{Rodriguez-Gomez}
  et~al.}{2016}]{Rodriguez-Gomez+2016}
{Rodriguez-Gomez} V.,  et~al., 2016, \mn@doi [\mnras] {10.1093/mnras/stw456},
  \href {http://adsabs.harvard.edu/abs/2016MNRAS.458.2371R} {458, 2371}

\bibitem[\protect\citeauthoryear{{Schaye} et~al.,}{{Schaye}
  et~al.}{2015}]{Schaye+2015}
{Schaye} J.,  et~al., 2015, \mn@doi [\mnras] {10.1093/mnras/stu2058}, \href
  {https://ui.adsabs.harvard.edu/abs/2015MNRAS.446..521S} {446, 521}

\bibitem[\protect\citeauthoryear{{Sijacki}, {Vogelsberger}, {Genel},
  {Springel}, {Torrey}, {Snyder}, {Nelson}  \& {Hernquist}}{{Sijacki}
  et~al.}{2015}]{Sijacki+2015:Illustris}
{Sijacki} D.,  {Vogelsberger} M.,  {Genel} S.,  {Springel} V.,  {Torrey} P.,
  {Snyder} G.~F.,  {Nelson} D.,   {Hernquist} L.,  2015, \mn@doi [\mnras]
  {10.1093/mnras/stv1340}, \href
  {https://ui.adsabs.harvard.edu/abs/2015MNRAS.452..575S} {452, 575}

\bibitem[\protect\citeauthoryear{{Singh}, {Mandelbaum}, {Seljak}, {Slosar}  \&
  {Vazquez Gonzalez}}{{Singh} et~al.}{2017}]{Singh+2017}
{Singh} S.,  {Mandelbaum} R.,  {Seljak} U.,  {Slosar} A.,   {Vazquez Gonzalez}
  J.,  2017, \mn@doi [\mnras] {10.1093/mnras/stx1828}, \href
  {http://adsabs.harvard.edu/abs/2017MNRAS.471.3827S} {471, 3827}

\bibitem[\protect\citeauthoryear{{Soko{\l}owska}, {Capelo}, {Fall}, {Mayer},
  {Shen}  \& {Bonoli}}{{Soko{\l}owska} et~al.}{2017}]{Sokolowska+2017}
{Soko{\l}owska} A.,  {Capelo} P.~R.,  {Fall} S.~M.,  {Mayer} L.,  {Shen} S.,
  {Bonoli} S.,  2017, \mn@doi [\apj] {10.3847/1538-4357/835/2/289}, \href
  {https://ui.adsabs.harvard.edu/abs/2017ApJ...835..289S} {835, 289}

\bibitem[\protect\citeauthoryear{{Speagle} et~al.,}{{Speagle}
  et~al.}{2019}]{Speagle+2019}
{Speagle} J.~S.,  et~al., 2019, arXiv e-prints, \href
  {https://ui.adsabs.harvard.edu/abs/2019arXiv190605876S} {p. arXiv:1906.05876}

\bibitem[\protect\citeauthoryear{{Springel} \& {Hernquist}}{{Springel} \&
  {Hernquist}}{2003a}]{Springel&Hernquist:2003a}
{Springel} V.,  {Hernquist} L.,  2003a, \mn@doi [\mnras]
  {10.1046/j.1365-8711.2003.06206.x}, \href
  {https://ui.adsabs.harvard.edu/abs/2003MNRAS.339..289S} {339, 289}

\bibitem[\protect\citeauthoryear{{Springel} \& {Hernquist}}{{Springel} \&
  {Hernquist}}{2003b}]{Springel&Hernquist:2003b}
{Springel} V.,  {Hernquist} L.,  2003b, \mn@doi [\mnras]
  {10.1046/j.1365-8711.2003.06207.x}, \href
  {https://ui.adsabs.harvard.edu/abs/2003MNRAS.339..312S} {339, 312}

\bibitem[\protect\citeauthoryear{{Springel} et~al.,}{{Springel}
  et~al.}{2018}]{Springel+2018:TNG}
{Springel} V.,  et~al., 2018, \mn@doi [\mnras] {10.1093/mnras/stx3304}, \href
  {https://ui.adsabs.harvard.edu/abs/2018MNRAS.475..676S} {475, 676}

\bibitem[\protect\citeauthoryear{{Stinson}, {Seth}, {Katz}, {Wadsley},
  {Governato}  \& {Quinn}}{{Stinson} et~al.}{2006}]{Stinson+2006}
{Stinson} G.,  {Seth} A.,  {Katz} N.,  {Wadsley} J.,  {Governato} F.,   {Quinn}
  T.,  2006, \mn@doi [\mnras] {10.1111/j.1365-2966.2006.11097.x}, \href
  {https://ui.adsabs.harvard.edu/abs/2006MNRAS.373.1074S} {373, 1074}

\bibitem[\protect\citeauthoryear{{Tacchella} et~al.,}{{Tacchella}
  et~al.}{2019}]{Tacchella+2019}
{Tacchella} S.,  et~al., 2019, \mn@doi [\mnras] {10.1093/mnras/stz1657}, \href
  {https://ui.adsabs.harvard.edu/abs/2019MNRAS.tmp.1573T} {}

\bibitem[\protect\citeauthoryear{{Teyssier}, {Moore}, {Martizzi}, {Dubois}  \&
  {Mayer}}{{Teyssier} et~al.}{2011}]{Teyssier+2011}
{Teyssier} R.,  {Moore} B.,  {Martizzi} D.,  {Dubois} Y.,   {Mayer} L.,  2011,
  \mn@doi [\mnras] {10.1111/j.1365-2966.2011.18399.x}, \href
  {https://ui.adsabs.harvard.edu/abs/2011MNRAS.414..195T} {414, 195}

\bibitem[\protect\citeauthoryear{{Torrey}, {Vogelsberger}, {Genel}, {Sijacki},
  {Springel}  \& {Hernquist}}{{Torrey} et~al.}{2014}]{Torrey+2014}
{Torrey} P.,  {Vogelsberger} M.,  {Genel} S.,  {Sijacki} D.,  {Springel} V.,
  {Hernquist} L.,  2014, \mn@doi [\mnras] {10.1093/mnras/stt2295}, \href
  {https://ui.adsabs.harvard.edu/abs/2014MNRAS.438.1985T} {438, 1985}

\bibitem[\protect\citeauthoryear{{Vogelsberger}, {Genel}, {Sijacki}, {Torrey},
  {Springel}  \& {Hernquist}}{{Vogelsberger} et~al.}{2013}]{Vogelsberger+2013}
{Vogelsberger} M.,  {Genel} S.,  {Sijacki} D.,  {Torrey} P.,  {Springel} V.,
  {Hernquist} L.,  2013, \mn@doi [\mnras] {10.1093/mnras/stt1789}, \href
  {https://ui.adsabs.harvard.edu/abs/2013MNRAS.436.3031V} {436, 3031}

\bibitem[\protect\citeauthoryear{{Vogelsberger} et~al.,}{{Vogelsberger}
  et~al.}{2014a}]{Vogelsberger+2014:Illustris}
{Vogelsberger} M.,  et~al., 2014a, \mn@doi [\mnras] {10.1093/mnras/stu1536},
  \href {http://adsabs.harvard.edu/abs/2014MNRAS.444.1518V} {444, 1518}

\bibitem[\protect\citeauthoryear{{Vogelsberger} et~al.,}{{Vogelsberger}
  et~al.}{2014b}]{Vogelsberger+2014:Nature}
{Vogelsberger} M.,  et~al., 2014b, \mn@doi [\nat] {10.1038/nature13316}, \href
  {https://ui.adsabs.harvard.edu/abs/2014Natur.509..177V} {509, 177}

\bibitem[\protect\citeauthoryear{{Wadsley}, {Stadel}  \& {Quinn}}{{Wadsley}
  et~al.}{2004}]{Wadsley+2004}
{Wadsley} J.~W.,  {Stadel} J.,   {Quinn} T.,  2004, \mn@doi [\na]
  {10.1016/j.newast.2003.08.004}, \href
  {https://ui.adsabs.harvard.edu/abs/2004NewA....9..137W} {9, 137}

\bibitem[\protect\citeauthoryear{{Wang} et~al.,}{{Wang}
  et~al.}{2019}]{Wang+2019}
{Wang} Y.,  et~al., 2019, \mn@doi [\mnras] {10.1093/mnras/stz2907}, \href
  {https://ui.adsabs.harvard.edu/abs/2019MNRAS.tmp.2518W} {p.~2518}

\bibitem[\protect\citeauthoryear{{Wang} et~al.,}{{Wang}
  et~al.}{2020}]{Wang+2018}
{Wang} Y.,  et~al., 2020, \mn@doi [\mnras] {10.1093/mnras/stz3348}, \href
  {https://ui.adsabs.harvard.edu/abs/2020MNRAS.491.5188W} {491, 5188}

\bibitem[\protect\citeauthoryear{{Weinberger} et~al.,}{{Weinberger}
  et~al.}{2017}]{Weinberger+2017}
{Weinberger} R.,  et~al., 2017, \mn@doi [\mnras] {10.1093/mnras/stw2944}, \href
  {https://ui.adsabs.harvard.edu/abs/2017MNRAS.465.3291W} {465, 3291}

\bibitem[\protect\citeauthoryear{{van den Bosch}}{{van den
  Bosch}}{2017}]{van-den-Bosch2017}
{van den Bosch} F.~C.,  2017, \mn@doi [\mnras] {10.1093/mnras/stx520}, \href
  {https://ui.adsabs.harvard.edu/abs/2017MNRAS.468..885V} {468, 885}

\bibitem[\protect\citeauthoryear{{van den Bosch} \& {Ogiya}}{{van den Bosch} \&
  {Ogiya}}{2018}]{van-den-Bosch+2018}
{van den Bosch} F.~C.,  {Ogiya} G.,  2018, \mn@doi [\mnras]
  {10.1093/mnras/sty084}, \href
  {https://ui.adsabs.harvard.edu/abs/2018MNRAS.475.4066V} {475, 4066}

\bibitem[\protect\citeauthoryear{{van den Bosch}, {Ogiya}, {Hahn}  \&
  {Burkert}}{{van den Bosch} et~al.}{2018}]{van-den-Bosch+2017}
{van den Bosch} F.~C.,  {Ogiya} G.,  {Hahn} O.,   {Burkert} A.,  2018, \mn@doi
  [\mnras] {10.1093/mnras/stx2956}, \href
  {https://ui.adsabs.harvard.edu/abs/2018MNRAS.474.3043V} {474, 3043}

\makeatother
\end{thebibliography}
\label{lastpage}


\end{document}